\documentclass[%
 reprint,
 amsmath,amssymb,
 aps,
]{revtex4-2}

\usepackage{graphicx}
\usepackage{dcolumn}
\usepackage{bm}
\usepackage{booktabs}

\usepackage{comment}
\usepackage{hyperref}

\begin{document}

\preprint{APS/123-QED}

\title[Diffuse probes]{High-resolution x-ray scanning with a diffuse, Huffman-patterned probe to minimise radiation damage}

\author{Alaleh Aminzadeh}
\author{Andrew M. Kingston}%
 \altaffiliation[Also at ]{CTLab: National Centre for Micro- Computed Tomography, Australian National University}%
\affiliation{ 
Department of Materials Physics, Research School of Physics, Australian National University, Australia}%

\author{Lindon Roberts}%
\affiliation{School of Mathematics and Statistics, University of Sydney, Australia}%

\author{David M. Paganin}%
\affiliation{School of Physics and Astronomy, Monash University, Australia}%

\author{Timothy C. Petersen}%
\affiliation{Monash Centre for Electron Microscopy, Monash University, Australia}%

\author{Imants D. Svalbe}%
 \email{imants.svalbe@monash.edu}
\affiliation{School of Physics and Astronomy, Monash University, Australia }%

\date{\today}

\begin{abstract}
Scanning objects with a more tightly focused beam (for example of photons or electrons) can provide higher-resolution images. However the stronger localisation of energy deposition can damage tissues in organic samples or may rearrange the chemical structure or physical properties of inorganic materials. Scanning an object with a broad beam can deliver an equivalent probe energy but spreads it over a much wider footprint. Sharp images can be reconstructed from the diffuse implanted signal when a decoding step can recover a delta-like impulse response. Huffman sequences, by design, have the optimal delta-like autocorrelation for aperiodic (non-cyclic) convolution and are well-conditioned. Here we adapt 1D Huffman sequences to design 2D Huffman-like discrete arrays that have spatially broad, relatively thin and uniform intensity profiles that retain excellent aperiodic autocorrelation metrics. Examples of broad shaped diffuse beams were developed for the case of x-ray imaging. A variety of masks were fabricated by the deposition of finely structured layers of tantalum on a silicon oxide wafer. The layers form a pattern of discrete pixels that modify the shape of an incident uniform beam of low energy x-rays as it passes through the mask. The intensity profiles of the x-ray beams after transmission through these masks were validated, first by acquiring direct-detector x-ray images of the masks, and second by raster scanning a pinhole over each mask pattern, pixel-by-pixel, collecting ``bucket'' signals as applied in traditional ghost imaging. The masks were then used to raster scan the shaped x-ray beam over several simple binary and ``gray'' test objects, again producing bucket signals, from which sharp reconstructed object images were obtained by deconvolving their bucket images.

\end{abstract}

\maketitle


\section{Introduction}
High resolution scanning x-ray imaging typically requires high fidelity beam shaping to create sharp probes \cite{SharpXrayProbeYan2014,SharpXrayProbeBajt2018}, concentrating the radiation that is rastered over the specimen. The same is true for atomic-resolution scanning transmission electron imaging, where it is advantageous to further remove probe aberrations in dedicated post-processing to achieve the highest resolution \cite{HighResEMMuller2021}. Irrespective of resolution, scanned electron beam shaping can be also useful for patterning the scattered radiation to sensitively measure strain \cite{STEMPatternProbesZELTMANN2020} or efficiently map crystal orientations \cite{MultBeamSTEMOphus2021}. An alternative to rastered x-ray scanning is to transmit a known ensemble of broad yet structured probes through a specimen, to be computationally reconstructed through ``ghost imaging'' \cite{GhostImagingReviewErkmen2010, GhostImagingPelliciaPRL2016, yu2016fourier}. Here we instead consider a broad yet detailed x-ray probe, encoding the sample through scanning, and seek to recover sharp images through correlative reconstruction.

This paper addresses the conceptual and experimental design, fabrication and application of 2D masks aimed at reducing the damage to specimens being scanned during x-ray imaging. A spatially broad beam of uniform intensity incident onto these masks transmits an equally broad but discretely shaped pattern of illumination. Each digital pattern is designed to have an autocorrelation that is, as close as possible, a delta function. Using this intensity pattern as an x-ray probe to scan a test object produces a blurred image of the object as it is convolved by the pattern of the mask. The inverse problem of decoding the pattern encoded by the mask is possible because of the mask's delta-like autocorrelation property. Prior work \cite{SvalbeTCI2020} addressed the design of diffuse arrays (here called Huffman-like arrays) that act as delta functions, whilst Ref. \cite{svalbe2021diffuse} looked at similar 2D mask patterns that are also delta-like when viewed as projections at multiple view angles.


In an experiment, the pattern of the broad shaped beam emerging from the mask is raster scanned across a test object. At each scan point, we integrate the intensity of x-rays transmitted through the mask that illuminates the mask-sized area of the test object. The integrated intensity forms the ``bucket signal'' for that scan point. The collection of all bucket signals obtained from the mask pattern placed over each raster scan point forms a ``bucket image''. The collection of bucket signals here is the same process used in ghost imaging \cite{kingston2018x,kingston2023optimizing}. The bucket image shows the test object convolved by the spatial pattern of the illuminating mask. Deconvolving the bucket image, by cross-correlating it with the discrete mask pattern, reconstructs a sharp 2D image of the scanned test object.

The motivation for imaging with a diffuse beam is to significantly reduce the impact on the scanned object of the amount and rate of energy deposited inside small regions of the test object. Energy deposition has paramount importance in the x-ray imaging of soft-tissues or nano-structured materials. Scanning an object with a diffuse beam having an area of $10\times10$ pixels, can, in principle, reduce the rate of incident energy deposition by $100$ times relative to that from an equivalent signal-to-noise incident beam focused to be one pixel wide.

Section \ref{sec:HuffmanDesign} reviews the theory behind the design of diffuse, delta-like pixellated masks that transmit a narrow range of stepped intensities using beams that project onto areas comprised of hundreds to thousands of pixels. The mask design was inspired by earlier work on 1D aperiodic sequences by Huffman \cite{Huffman1962}. Section \ref{sec:X-ray imaging using 2D Huffman-like beams} describes the design and the challenging techniques developed here to fabricate Huffman-like masks; this includes both binary masks suited for use with x-rays with an energy of around 20 keV, and quaternary masks designed for use with x-ray energy of 12.4 keV. The latter masks were built as multiple uniform layers of tantalum deposited in discrete patterns on a silicon oxide wafer. Section \ref{sec: Mask Validation} presents experimental results obtained using a uniform beam of synchrotron generated x-rays that was shaped by transmission through these masks. Validation of the as-fabricated patterns was achieved by directly imaging the x-ray beam transmitted by each mask. Several simple binary objects and more complex ``gray'' test objects were scanned with these diffuse beams to produce bucket images from which images were reconstructed that closely match the test objects. We discuss the experiment results and their implications in Sec. \ref{sec:discussion}. Section \ref{sec:conclusion} summarises our findings and discusses future practical imaging applications for broad Huffman-like masks. Supplementary Material is provided to give additional relevant information on the properties of Huffman sequences, the design and testing of compressed Huffman-like arrays and a more detailed overview of the fabrication design and wafer production process necessary to make the masks used in the x-ray imaging experiment.

\section{Numerical design of broad, diffuse x-ray probes with delta-like auto-correlation} \label{sec:HuffmanDesign}

Scanning probes acquire images of test objects that are blurred by the spread and shape of the scanning probe. 
To reduce the impact of the local radiation dose rate of the incident beam, the probe area could be made broader or more diffuse. A broader beam generally means poorer spatial resolution. However a diffuse probe shape that is encoded with a pattern that has an autocorrelation closely approximating a delta function can circumvent this issue; we refer such probes as having {\it delta-like autocorrelation}. The object image that was blurred by the shape of the probe can be restored by a deconvolution using the known shape of the probe. Probes that have delta-like autocorrelation are well-conditioned and ensure that the inverse deblurring problem is robust to measurement noise. The method may be viewed as a form of coded-aperture imaging \cite{Coded-aperture-imaging-review}, in which the coding and decoding arrays are identical to one another.

This section describes how to encode such sequences into suitable shapes for use as scanned x-ray probes. Much attention has been given in the literature to the many and varied periodic sequences that are delta-like under circular convolution or correlation (so-called {\it periodic arrays}) \cite{Chu1972,SchmidtPerfectArrays,LegendreSvalbePaganinCavyPetersen2024}. However, these sequences are impractical when the probe field-of-view (FOV) limits the object size. We usually need to scan objects larger in size than the probe FOV. In this situation, the probe performs non-circular (or aperiodic) convolution or correlation. We hence seek to encode (or pattern) the probe shape using sequences that possess delta-like \emph{aperiodic} autocorrelation. The Huffman sequences described in the next section are, by design, optimal for aperiodic convolution. 

\subsection{Huffman Sequences and 2D Huffman Arrays}\label{sec:HuffmanDefinition}

In 1962, Huffman
in a pioneering paper \cite{Huffman1962}, defined what constitutes the optimal possible delta-like form for any aperiodic autocorrelation and then constructed quite special examples of sequences, $H_L$, for any length $L$, that met that goal.

Following that work, several different types of integer, real, and complex Huffman sequences have been found. These sequences, specified here as $H_L^s$ for distinct sequence types $s$, all have in common the following unique property under \emph{aperiodic} autocorrelation $(\otimes)$:
\begin{equation}\label{eq:HuffDefinition}
    H^s_L \otimes H^s_L \approx \delta.
\end{equation}
Here $\delta$ is the Kronecker delta function:
\begin{equation}\label{eq:deltaDefinition}
    \delta(x-a) = 0, x\ne a,
\end{equation}
with $x$ and $a$ being array coordinates.
 So-called ``perfect-sequences'', $S_L$, where 

\begin{equation}\label{eq:PerfectDeltaDefinition}
    S_L \otimes S_L \equiv \delta
\end{equation}
under \emph{periodic} autocorrelation conditions, were well-known long before Huffman's work \cite{SchmidtPerfectArrays}. However those sequences do not satisfy Eq.~(\ref{eq:HuffDefinition}) when used in aperiodic operations.

Building on Huffman's paper, Hunt and Ackroyd \cite{HuntAckroyd1980}, discovered integer-valued Huffman sequences (of length $L = 4n-1$) with entries that turn out to be derived from the Lucas/Fibonacci series. An example of their sequence type is 

$$H_7 = [1, 2, 2, 0, -2, 2, -1].$$
The aperiodic autocorrelation of $H_7$ is:

$$H_7 \otimes H_7 = [-1, 0, 0, 0, 0, 0, 18, 0, 0, 0, 0, 0, -1].$$

An aperiodic autocorrelation is \emph{optimally delta-like} (following Eq.~(\ref{eq:HuffDefinition})), when all of the off-peak autocorrelation values are zero, bar the unavoidably non-zero left and right end correlation values. The magnitude of the end values are kept as small as possible relative to the autocorrelation peak, which should be as large as possible. For $H_7$, those values are $-1$ and $18$, respectively. ``Perfect-periodic'' sequences, for example Legendre $L_7 = [0,1,1,-1,1,-1,-1]$, are not delta-like (for $L_7$, the aperiodic autocorrelation peak is $6$ with many non-zero off-peak values ranging from $+1$ to $-2$). Thus

$$L_7 \otimes L_7 = [0,-1,-2,1,0,-1,6,-1,0,1,-2,-1,0].$$

The range of and spacing between the values in the elements of $H_L^s$ has critical practical implications for the arrays fabricated in our experiment, for example to construct the $32\times32$ array. The $1D$ Lucas/Fibonacci sequence $H_{31}$, has signed integer values that range between $\pm{754}$. The integer range for these Huffman sequences grows exponentially with length $L$:  
    
\begin{equation}\label{eq:Huffman max range}
max(|H_L|) = \lfloor (2/\sqrt{5})\times\phi^{(L-3)/2} \rceil
\end{equation}
where $\phi$ is the golden ratio: $(1+\sqrt{5})/2$, and $\lfloor r \rceil$ denotes the integer round operation on real value $r$. The ranges fixed by Eq.~(\ref{eq:Huffman max range}) motivate the compression of Huffman sequence values, as covered in Sec. \ref{sec:HuffmanCompression}.

In general, any \emph{canonical} Huffman sequence of length $L$ has optimal aperiodic autocorrelation of length $2L-1$ with the form $$[lr,0,0,\cdots,0,A_0,0,\cdots,0,0,rl].$$
The peak of the auto-correlation value at zero shift is $A_0$, whilst the (unavoidable) end terms $rl= lr$ arise from the product of the (non-zero) leftmost and rightmost elements of the sequence $H^s_L$, denoted here as $l$ and $r$, respectively. Then Eq.~(\ref{eq:HuffDefinition}) is the closest possible approximation to an aperiodic $\delta$, especially for integer sequences when $rl = \pm1$ (although $|l|$ need not match $|r|$ for all Huffman sequence types).
 
The elements of Huffman sequence $H_L$ are comprised of $L$ signed values that, under the convolution operator, act as a discrete, broadband filter. To be delta-like, the bandwidth of $H_L^s$ as a filter has to have a near-constant response over all $L$ spatial frequencies in order for Eq.~(\ref{eq:HuffDefinition}) to hold. Further examples of integer Huffman sequences and their properties are given in the Supplementary Material, together with a detailed description of their Fourier spectra. 

Huffman derived the necessary and sufficient criteria to compute canonical sequences $H^s_L$ in terms of a complex polynomial $P(z)$: in the complex plane, any root/zero $z_l$ of $P(z)$ must lie on a circle of radius either $R$ or $1/R$ with phase angle $\textrm{Arg}(z_l) = 2\pi l/(L-1)$. See Ref.~\cite{Huffman1962} and related work in Refs.~\cite{Ackroyd1977} and \cite{OjedaTacconiFlatHuffmans}. 

A polynomial with $L$ coefficients $c_l$ where
$$\sum c_l z^l \equiv c_{L-1}\prod (z-z_l),$$
can be represented as a discrete Fourier transform $\mathcal{F}$ when evaluated on the unit-circle by setting $z = \exp(2\pi i q/L)$, where the integer $q$ lies in the range $0 \le q \le (L-1)$. Given a set of roots $z_l$, an inverse Fourier transform $\mathcal{F}^{-1}$ can thus efficiently compute and sort the coefficients $c_l$.

Using Huffman's criteria for the roots placement, the $q^{th}$ Fourier coefficient $\mathcal{F}[H_{L}^s]_q$ of any canonical Huffman sequence hence can be calculated from,
\begin{equation}\label{eq:canonicalHuffmanCompute}
\mathcal{F}[H_{L}^s]_q =  c_{L-1}\prod_{l=1}^{L-1} (e^{2\pi i q/L}-R^{s_l}e^{2\pi i (l-1)/L}),
\end{equation}
where $s_l \in \{-1,+1\}$ is any chosen set of signs. While the elements of $H_{L}^s$ are sensitive to the $2^{L-1}$ choices for $s_l$, each choice produces an identical canonical autocorrelation function. How Eq.~(\ref{eq:canonicalHuffmanCompute}) adheres to Huffman's original construction is explained in the Supplementary Material, as is the straightforward restriction to real-valued $H_{L}^s$. By careful choice of $R$, the growth of $max(|H_{L}^s|)$ with increasing $L$ (such as in Eq.~(\ref{eq:Huffman max range})) can be suppressed, yet the values are generally non-integer. Therefore the dynamic range of $H_{L}^s$ still increases for larger $L$.

Huffman sequences that have a  compact range of element values and close-to-zero off-peak aperiodic autocorrelation elements are particularly suited for x-ray imaging applications. We call sequences that are optimised to adhere closely to Eq.~(\ref{eq:HuffDefinition}) ``Huffman-like''.

\subsection{Encoding x-ray probes with 2D Huffman arrays} \label{sec:Huffman2Dxray}
A patterned scanning probe based on a 2D Huffman array $H$ can be constructed from 1D $H^s$ sequences using an outer product $*$, such that $H = H^s*H^s$ also obeys the desired property given in Eq.~(\ref{eq:HuffDefinition}) \cite{SvalbeTCI2020}. Suppose a 2D x-ray probe has been encoded by a mask defined by $H$. For an object of interest $O$ convolved ($\circledast$) with $H$, a sharp image of $O$ can be obtained by cross-correlation via property Eq.~(\ref{eq:HuffDefinition}):  
\begin{equation}\label{eq:deblurHuffman}
    O \approx [O \circledast H] \otimes H.
\end{equation}
%




Thus far we have described how to generate 2D Huffman arrays with the optimal aperiodic autocorrelation property. The delta-like autocorrelation means that we can robustly compute a high-resolution image (of arbitrary size due to the aperiodic probe properties), by scanning a test object with a broad beam patterned by these arrays. In the context of x-ray imaging, a beam with a wider footprint is often preferable, to diffuse the x-ray dose more effectively. However, as shown by Eq.~(\ref{eq:Huffman max range}), the range of array values grows rapidly with increased array size. The fabrication of masks that are able to transmit x-rays with a large range of discrete intensity values (or a wide range of discrete phase shifts in the wavelength) is intrinsically difficult. Practical construction concerns require we place strong limits on the value range of transmitted x-ray intensities (or phase shifts).

Experimentally, a physical attenuation used to moderate x-ray intensities can only be comprised of non-negative variations in intensity. We propose a method to achieve this in the following section (Sec. \ref{sec:masksHuffmanPN}). To be practical from a fabrication point of view, the mask should also be restricted to a providing only a few different intensities comprised of clearly-separated uniform steps. This is achieved by compressing the dynamic range of Huffman sequences to manageable levels whilst still retaining their Huffman properties. The quality metrics used to quantify array performance are detailed in Sec. \ref{sec:metricsHuffman}. Having established these metrics, we then present two different design approaches in Sec. \ref{sec:HuffmanCompression} to ensure that Eq.~(\ref{eq:HuffDefinition}) is maintained, as closely as is possible, after compressing the gray-level range of Huffman arrays. 

\subsubsection{Non-negative 2D X-ray Masks to Imprint Huffman Arrays}\label{sec:masksHuffmanPN}

Huffman arrays necessarily comprise signed elements, while x-ray intensities are non-negative. Huffman arrays can nonetheless be realised using pairs of absorptive x-ray masks, provided that the decorrelation method Eq.~(\ref{eq:deblurHuffman}) is generalised. We split the signed values of the Huffman-like array $H$ into two positive-definite parts as ``Huffman masks'' - one comprising the positively signed elements, $P = max[H, 0]$, and another containing the magnitudes of the negative elements, $N = max[-H, 0]$, such that
\begin{equation}\label{splitDiff}
H = P-N.     
\end{equation}
If a signal $S_T$ encodes an object $O$ of interest with a Huffman-like array via a linear relation such as $S_T=O\circledast H$, this total signal can be decomposed as the difference between two successive measurements $S_p$ and $S_n$ using the Huffman mask arrays $P$ and $N$ respectively,
\begin{equation}\label{eq:decorrSplit}
S_T =S_p-S_n=O \circledast P-O \circledast N = O\circledast H.
\end{equation}
The desired object signal $O$ can then be de-correlated by cross-correlating $S_T$ with the delta-correlated $H$.

In our experiment, the x-ray beam, after being shaped by the array, is to be raster scanned across the object being imaged. Both $P$ and $N$ masks need to be applied separately. We arranged each $P$ and $N$ mask in a $2\times2$ block, i.e.,
\begin{equation*}
\begin{bmatrix}
    P & N \\
    N & P
\end{bmatrix},
\end{equation*}
that we will denote inline as $[P,N/N,P]$. This layout permits the $P$ and $N$ probes to be scanned separately across the object, along either the horizontal or vertical scan axes. This redundant mask arrangement also provides multiple ways to combine $P$ and $N$ bucket images to form the bucket image from any mask $H$.

Having solved how, in practice, to deal with negative mask intensities, the next problem is to constrain the number of intensity levels each $P$ and $N$ mask transmits. More and more finely-spaced transmission levels impose significantly tighter demands on mask fabrication. The range of Huffman mask values can be compressed, but practical truncation or rounding of the compressed levels rapidly reduces their essential delta-like autocorrelation property.

We chose to limit the design range of our Huffman-like arrays to $\pm3$. The $P$ and $N$ arrays then have four uniformly spaced steps: $[0,1,2,3]$, where $0$ means the lowest and $3$ means the highest transmitted intensity. An intensity range larger than $\pm3$ would prove, in practice, difficult to fabricate for x-rays. A key aim of this experiment is to keep the broad beam intensity profile as wide and uniform as possible to minimize the rate of energy deposition.

For larger sized arrays, we found that limiting arrays values to $\pm2$ produced a mostly binary $\pm1$ result that cannot capture the broad central peak of intensities that is typical of classic integer Huffman sequences. On the other hand, each array element that has value $\pm3$ increases the size of the autocorrelation peak $A_0$ by $9$ (and each $\pm2$ value adds increases of $4$). In contrast, array values of $\pm1$ add just $1$ to $A_0$. Sequences with all values $\pm1$ are known to not satisfy Eq.~(\ref{eq:HuffDefinition}), which is essential for our imaging context (excluding the special case of area-weighted binary masks presented in Sec. \ref{sec:HuffmanBinarising}).

The relevant metrics used to guide the range compression process are described in the next section.

\subsubsection{Array Delta-correlation Quality Metrics}\label{sec:metricsHuffman}
Here we present metrics to quantify the desirable properties of the diffuse probe patterns. All the following metrics were monitored and simultaneously optimised when compressing Huffman array element values. The metric definitions are shown below for sums over the $2D$ arrays used here, but apply equally for $1D$ and $nD$ arrays, where the sums are taken over all array elements. 

Our primary aim is to minimise the adverse effects of the probe causing local heating or radiation damage to the scanned object. To achieve this, the diffuse x-ray probe should transmit close-to-uniform intensities over the entire area of the mask to maximally spread the dose. This property can be quantified for an $L \times L$ array $H_L$ by:
\begin{description}
    \item[RMS] The root mean square of intensities transmitted by an array is:
    $$RMS = \sqrt{\frac{1}{L^2}\sum_i\sum_j H_L(i,j)^2},$$
    
    RMS values of 1 correspond to the same uniform (positive) intensity transmitted through each array element.
    
    \item[MAV] The mean absolute value is defined as:
    $$MAV = \frac{1}{L^2}\sum_i\sum_j |H_L(i,j)|.$$
    MAV values of 1 (for uniform transmission across the mask) are preferred.
\end{description}

An array with a strongly compressed range of integer intensities may also set many array elements to zero, reducing the RMS and MAV values closer to 1. However, opaque regions in an x-ray mask do not transmit specimen information and so diminish the imaging efficiency. A further useful quality measure is then:
\begin{description}
    \item[$\mathbf{f^z}$] the fraction $f^z$ of zero intensities in the array pattern,
    $$f^z = (1/L^2)\sum_i\sum_j \delta(H_L(i,j)),$$
    where $\delta(\eta) = 1$ when $\eta = 0$ and 0 otherwise. Arrays with $f^z \approx 0$ (with any zero elements located peripherally rather than centrally) are preferred. This metric can also be applied to the $(2L-1)\times(2L-1)$ autocorrelation array. Then more zeros are preferred and $f^z \approx 1$.
\end{description}

In this work we use three different measures to quantify how closely the autocorrelation of each array is delta-like to satisfy  Eq.~(\ref{eq:deblurHuffman}). Here the array peak autocorrelation value is $A_0$ and the off-peak autocorrelation values are $A_{i,j}$, for shifts $-L \le (i,j) \le L$, for $(i,j)$ not both zero.
\begin{description}
    \item[$\mathbf{M^f}$] The ``merit factor'' is the square of the autocorrelation peak value divided by the sum of all squared off-peak autocorrelation values. Here larger values are preferred, with
    $$M^f = A_0^2 / \sum A_{i,j}^2. $$
    \item[$\mathbf{R^o}$] The ``peak to side-lobe'' is the ratio of the peak autocorrelation value to the largest of all the off-peak autocorrelation absolute values. Larger values are preferred, with
    $$R^0 = A_0/max(|A_{i,j}|).$$
    \item[$\mathbf{d^F}$] ``Spectral flatness'' has several possible definitions, one being the ratio of the geometric to arithmetic mean of the Fourier magnitudes (as used in acoustics). Here we use:
    $$d^F = [\max(|\mathcal{F}[H]|) - \min(|\mathcal{F}[H]|)]/\mathrm{mean}(|\mathcal{F}[H]|),$$
    as this measure proved to be more sensitive to the extremes of variation in the magnitude of the Huffman-like array Fourier transform coefficients $|\mathcal{F}[H]|$ . Values closer to zero are preferred.
\end{description}

Well conditioned arrays are preferable since they are more robust to noise and inversion is a stable process. To measure how well conditioned the arrays are (especially after range compression), we use the singular value decomposition (SVD) of the array, taken under zero-padded aperiodic boundary conditions, with singular values stored in the set $\sigma(H_L)$.
\begin{description}
    \item[$\mathbf{\kappa}$] The condition number is defined as the ratio of largest to smallest singular values, i.e,
    $$\kappa = max(\sigma(H_L))/min(\sigma(H_L)).$$
    Note that $\kappa \ge 1$ and values closer to one are preferred.
\end{description} 
%

We next compare example metrics for four $11 \times 11$ arrays. The first 2D array is built from the 1D integer Huffman sequence $H_{11}$. The second array, $H^c_{11}$, is a compressed integer range Huffman-like version built from the first array. The third is a 2D $B_{11}$ array, built from a length $11$ binary Barker sequence using an outer product. Barker sequences have unit-magnitude integer elements (i.e. ``binary''), with aperiodic auto-correlation values all $\le 1$. The longest known Barker sequence has 13 elements. Infinite families of Barker-like sequences with merit factors $M^f > 6.34$ can be constructed \cite{BarkerBorweinJedwabMeritFactor2004}. Recent stochastic algorithms have been designed to discover and optimise new such sequences \cite{BoskovicBarkerLike2017, BoskovicBarkerLike2024}.
The fourth is $L_{11}$, built from the length $11$ binary periodic ``perfect'' Legendre sequence \cite{GolayLegendre,HolholdtLegendre,LegendreSvalbePaganinCavyPetersen2024}, circularly shifted to maximise the aperiodic metrics $R^0, M^f$. 







\begin{center}
\begin{tabular}{|c|c|c|c|c|c|c|c|c|} 
 \hline 
 type & range & RMS & MAV & $f^z$ & $M^f$ & $R^o$  & $d^F$ & $\kappa$\\ [0.5ex] 
 \hline\hline
 $H_{11}$ & $\pm36$ & 11.18 & 7.94 & 0/121 & 3782 & 123.0 & 0.033 & 1.02 \\ 
 \hline
 $H^c_{11}$ & $\pm3$ & 1.52 & 1.30 & 11/121 & 18.24 & 35.13 & 0.635 & 1.31 \\
 \hline
 $B_{11}$ & $\pm1$ & 1.00 & 1.00 & 0/121 & 5.81 & 11.00 & 1.26 & 2.00 \\
 \hline
 $L_{11}$ & $\pm1$ & 0.91 & 0.83 & 21/121 & 2.55 & 5.00 & 2.09 & 4.49 \\
 \hline
\end{tabular}
\end{center}

These examples demonstrate that, under aperiodic convolution, Huffman arrays are indeed very strongly delta-like. This remains true for their Huffman-like versions, even after their integer element values have been compressed (from $\pm36$) to a more practical range ($\pm3$). In contrast, the Barker binary array shows poorer autocorrelation metrics relative to the Huffman-like array, as does the Legendre binary ``perfect'' array (the latter being designed to excel in the periodic domain). The autocorrelation metric values $R^0, M^f$ generally \emph{increase} with array size (as more element values are added) however it becomes more difficult to keep $d^F$, the Fourier flatness metric, closer to zero.

\subsection{Compressing Huffman Arrays as Practical Masks to Encode Diffuse Probes }\label{sec:HuffmanCompression}

The dynamic range of Huffman array elements can, only to some extent, be controlled by choice of the chosen phase distribution. Similarly the radius and sign choices, ($R$ and $s_l$ respectively in Eq.~(\ref{eq:canonicalHuffmanCompute})), can reduce extrema in Huffman-like arrays. To improve upon such initial choices, systematic procedures are developed here to further compress Huffman arrays while optimising the metrics of Sec.~\ref{sec:metricsHuffman}. We refer to compressed integer-valued Huffman element values as Huffman-like arrays having a small range of (signed) integer image gray levels.

\subsubsection{Iterative Optimisation of Compressed Huffman Array Element Values} \label{sec:HuffmanImantsCompression}

For integer Huffman sequences, the range of values usually grows with length, as given by Eq.~(\ref{eq:Huffman max range}) for the example of the Lucas/Fibonacci sequences. Recall that for $H_{31}$ the integers range over $\pm754$ and that the value range for a $2D$ array is the square of the 1D range. For practical mask fabrication, where constructing as few levels as possible is preferred, strong range compression is mandated, especially for arrays with side-lengths $L>{7}$.

Scaling each Huffman element value by the same constant preserves all autocorrelation metrics. The non-linear operation of integer rounding to enact range-compression on the other hand alters the otherwise flat power spectrum, which can in turn degrade all correlation metrics. To counter this, we iteratively optimise the metrics of Sec.~\ref{sec:HuffmanDesign} by initially down-scaling the Huffman array such that the smallest element values retain unit-magnitude (rather than rounding element values to zero). All element values are scaled down by a real number $v$ (by no more than a factor of 2) in fine steps while the Sec.~\ref{sec:HuffmanDefinition} metrics are monitored after integer rounding. When an optimal value of $v$ is determined, ``dithering'' is used by adding an $L \times L$ array of zero-mean white noise, with maximum magnitude less than 1/2. The Sec.~\ref{sec:HuffmanDesign} metrics are monitored after integer rounding, again to find an optimal random perturbation. The process of down-scaling and optimizing arrays is iterated until the desired dynamic range (of $\pm3$) is reached. 

The sequence compression process is driven by monitoring for continual improvement in the metrics $R^0, M^f$ and $d^f$, usually in that order of priority (as randomly structured sequences can be spectrally flat). When conflict in metric gains occurs, such as $R^0$ decreasing while $M^f$ improves, the compression step size is reduced and metric $d^f$ is given more priority. 

The final optimisation step randomly cycles through each array value (for example choosing those array elements with value $-2$) and perturbs a small random sample of those elements by changes of $\pm 1$, again monitoring for changes in the Sec.~\ref{sec:HuffmanDesign} metrics to fine-tune the array.

After random perturbation of $2D$ arrays, transpose symmetry is restored by taking (and rounding) the mean of the array and its transpose after each optimisation step. However, while our preference is to preserve this symmetry, it is possible that the transpose operation produces an array with an inferior set of metrics. In this case the asymmetric array is retained. 

Whilst outer products of a canonical Huffman sequence of length $L$ are suitable for building small $L \times L$ arrays, for longer lengths, (i.e., $L > 100$), compression of very large dynamic ranges (as given by Eq.~(\ref{eq:Huffman max range})) can yield erratic noise-like sequences that are difficult to optimise. Airy functions can instead be used as flatter-ranged impulse-like sequences, or ``chirps''. Portions of these $1D$ chirps are used to construct larger Huffman-like $2D$ arrays via an outer-product (an example of a range-compressed $120\times120$ array is given in the Supplementary Material).

\subsubsection{Hybrid Monte Carlo Optimisation of Huffman Array Element Values}\label{sec:HuffmanMonteCarlo}

Bernasconi \cite{BernasconiMonteCarlo1987} showed that maximizing the merit factor $M^f$ of binary sequences is akin to minimising the energy of spins in a $1D$ Ising model. Bernasconi's $1D$ Ising approach for constructing Barker-like binary sequences \cite{BarkerBorweinJedwabMeritFactor2004} (for example $B_{11}$ as given earlier) using Metropolis-Hasting Monte Carlo simulations \cite{BernasconiMonteCarlo1987} needs to be generalised here to construct integer-valued arrays with autocorrelation adhering to Eq.~(\ref{eq:deblurHuffman}).

To implement Markov steps, with reference to Eq.~(\ref{eq:canonicalHuffmanCompute}), we choose a starting radius $R$ to define an initial Huffman sequence, and a random set of signs $\{s_1,s_2,...s_{L-1}\}$ is fixed. A sign $s_c$ is then randomly chosen and randomly incremented/decremented by a small (sub-integer) random step size, sampling from a uniform distribution, such that the magnitudes of the complex roots are now defined by the non-canonical set $\{R^{s_1}, R^{s_2}, ... , R^{s_c}, ... R^{s_{L-1}}\}$ (where $|s_c|$ now differs from unity). After quantizing the associated Huffman sequence elements (or array for 2D), this random non-integer $s_c$ change is then accepted outright if the merit factor improves or conditionally accepted according to an exponential distribution of the weighted square difference. 

Our method for iteratively evolving the complex root radii was tested here for Barker-like unit-magnitude 1D sequences, while also mimicking simulated annealing by linearly varying the square-difference weight for the change in merit factor, in addition to the step size (to optimise the ratio of accept/reject moves). All known Barker sequences were efficiently found from these simulations in this work, in addition to all Barker-like sequences with largest possible merit factor (tested up to length 20). 

This generalization of Bernasconi's energy based Monte Carlo algorithm \cite{BernasconiMonteCarlo1987} only optimises the merit factor $M^f$, yet we need to optimise all metrics of Sec.~\ref{sec:metricsHuffman} in order to achieve the desired property in Eq.~(\ref{eq:HuffDefinition}) for Huffman-like arrays. With further adaption here, hybrid reverse Monte Carlo \cite{RMCMcGreevy1988, HRMCOpletal2001} (HRMC) is ideally suited for this purpose. While originally designed to minimise energy and maximise consistency with experimental diffraction measurements in atomic systems, including larger scale porosity constraints \cite{HRMCPorosityManifolds, HRMCPorosityOpletal}, HRMC optimisation can be used in radically different contexts (cf. a recent urban design study of flooding \cite{HRMCUrbanDesignBalaian2024}). 

For such optimisation of Huffman arrays in 2D, the quantization we have chosen here is
\begin{align}
\begin{split}
H_{2D quant} = \lceil (H-\langle H \rangle)*(H-\langle H \rangle) g/H_m^2 \rfloor,
\end{split}
\label{eq:quantise2DHuffman}
\end{align}
where $\langle H \rangle$ is the mean and $H_m = max{|H|}$ is the maximum of the statistically evolving $1D$ sequence, and $g$ is the maximum gray value magnitude. 

In this work, the connection to HRMC is quite natural if the Fourier spectrum $|\mathcal F[H_{2D quant}]|^2$ is viewed as a diffraction pattern, which ought to be constant across all spatial frequencies to ensure that $d^F$ of Sec.~\ref{sec:metricsHuffman} is optimised. We have thereby modified the HRMC algorithm to incorporate all of the metrics $M^f, R^o, f^z, d^F$ etc. of Sec.~\ref{sec:metricsHuffman}, with the exception of $\kappa$ due to the significant computational complexity of SVD. The adapted HRMC algorithm here minimises a $\chi^2$ computed from the Eq.~(\ref{eq:quantise2DHuffman}) array, relative to desired target values of all metrics. 
%
%

A trial move of a randomly chosen radius power's fractional sign $s_c$ is accepted if $\chi^2$ goes down, otherwise a random number $r$ on $[0, 1]$ is chosen and compared to $\exp[-(\chi^2_{trial} - \chi^2_{previous})]/(kT)$ for conditional acceptance/rejection if $r$ is smaller/larger respectively, where $kT$ is a global fictitious temperature. This differs slightly from conventional HRMC, whereby the temperature pertains only to the energy term. The purpose of the temperature here is to allow for simpler ``simulated annealing'' optimisations, whereby all weighting factors can be gradually reduced (as done individually in conventional HRMC) to improve numerical efficiency. Similarly, the random step size of the randomly-walked complex polynomial zero radii in Eq.~(\ref{eq:canonicalHuffmanCompute}) can be weighted by $kT$ here to maintain an efficient accept/reject ratio for Monte Carlo trials. Further computational efficiency is achieved by gradually compressing the dynamic range $g$ in Eq.~(\ref{eq:quantise2DHuffman}) (similar to the iterative methods in Sec.~\ref{sec:HuffmanImantsCompression}). 

Further details appear in the Supplementary Material, in addition to a simplified HRMC simulation example for creating a 2D Huffman-like array. 


\subsubsection{Reformatting Huffman Arrays as Binary Area-weighted Masks}\label{sec:HuffmanBinarising}

We forecast that the practical fabrication of masks with microscopic feature size and multiple uniform steps of transmitted x-ray intensity would be technically challenging. As a precautionary measure, we designed equivalent Huffman-like masks that had binary intensity transmission. These masks emulate multiple intensity levels by depositing absorbing material to cover fixed fractions of the pixel area.

For diffuse x-ray probes defined by Huffman-like array masks, one can sacrifice the spatial resolution of array elements in order to spread a given integer value across an effective ``sub-pixel'' (or ``sub-voxel'' for $nD$ arrays), which we will refer to as a sub-element array. We have implemented an entirely binary approach here which preserves the auto-correlation behaviour of the range-compressed and optimised Huffman-like arrays of Sec.~\ref{sec:HuffmanImantsCompression} and Sec.~\ref{sec:HuffmanMonteCarlo}. 

For a given element of $P$ or $N$ in Sec.~\ref{sec:masksHuffmanPN}, the non-negative integer value can be decomposed as a binarised sub-element array $Se$ comprising values either zero or unity, which we assume to be square for simplicity. We refer to a correlating/convolving sub element array as $Se_c$ and a given de-correlating array as $Se_d$. 

To preserve the cross-correlation of $H = P-N$, the inner product between any pair of sub-element arrays $Se_c$, $Se_d$ must match the products between pairs of elements of $H$. Denoting a sequence $b_s$ of zeros and ones, a simple choice is the outer product $Se_c = b_c\boldsymbol{1}^\intercal$ and $Se_d = \boldsymbol{1} b_d^\intercal$, where $\boldsymbol{1}$ is a sequence of unit-valued elements (an identity array for generalisation to binarized sub-voxels). In other words, $Se_d$ is geometrically orthogonal to $Se_c$, such that their inner product is simply the number of ones in $b_c$ times that in $b_d$. For arrays with transpose symmetry, the encoding and decorrelation scheme of Eq.~(\ref{eq:decorrSplit}) remains valid. A more general scheme that need not assume this symmetry is described in the Supplementary Material.   

The aperiodic auto-correlations $H\otimes H$ and $(B_p-B_n) \otimes (B_p-B_n)$ are exactly the same when the latter is sampled over element shifts, with sub-element shifts deemed a form of oversampling. To ensure precise integral sub-divisions, the binarised arrays $B_p$ and $B_n$ need be $(max{|H|})^n$ times larger than $H$, where $n$ is the number of dimensions. For applications where the array size is a limiting factor, this increase emphasises another reason why it is important to control the dynamic range of Huffman gray levels, $max(|H|)$. 

When choosing how to subdivide binarised sub-element arrays, it is important to avoid correlated placements that could impair the delta-like performance of the entire Huffman array for sub-element shifts. As such, we have implemented random partitioning of any sub-element array into opaque and transmission sub-elements (i.e. $b_1$, $b_2$ generated randomly). However this randomised sub-structure (or many other binary placements) can be quite detailed, which implies impractical synthesis for physical manufacturing of entire $B_p$ and $B_n$ arrays. For the x-ray masks, we have hence implemented an alternative blocked-design, whereby sub-elements of equal value are grouped together as separate blocks of either zero or one, the binary ordering of which is randomly chosen within the sub-element array. Explicit numeric examples of $2D$ sub-element arrays and a binarised Huffman-like array are visualised in the Supplementary Material. 

\section{X-ray scanning probe mask fabrication}
\label{sec:X-ray imaging using 2D Huffman-like beams}
\label{sec:mask_fab}

Given that we now have methods to design masks as diffuse probes with delta-like autocorrelation, we would like to evaluate the concept experimentally with x rays. We designed Huffman-like arrays to be constructed as physical x-ray attenuation masks with array sizes ranging from $11 \times 11$ to $86 \times 86$ and seven transmitted intensity values, (i.e., integer values ranging from $-3$ to $+3$ where 3 indicates transmission as close to $100\%$ as physically achievable). Since negative transmission cannot be achieved, the values in each array were separated into positive ($P$) and negative ($N$) masks (as described in Sec. \ref{sec:masksHuffmanPN}) each with four equally spaced transmitted intensities. 

The physical masks were fabricated on silica ($\textrm{SiO}_2$) wafers. This was imprinted with multiple layers of tantalum (Ta) through a combination of lithography and etching techniques. Several instances of each mask design were fabricated with array element (or ``pixel'') sizes in the mask patterns varying from 8 to 20 $\mu$m. The masks fabricated with four intensity levels will be referred to as {\it quaternary masks}. Binary versions of these masks were also designed (as described in Sec. \ref{sec: Binary Huffman-like Masks}) with each mask ``pixel'' divided into $3 \times 3$ subpixels. The masks fabricated with two intensity levels will be referred to as {\it binary masks}.

\begin{figure}[h!]
    \centering
    \includegraphics[width=\linewidth]{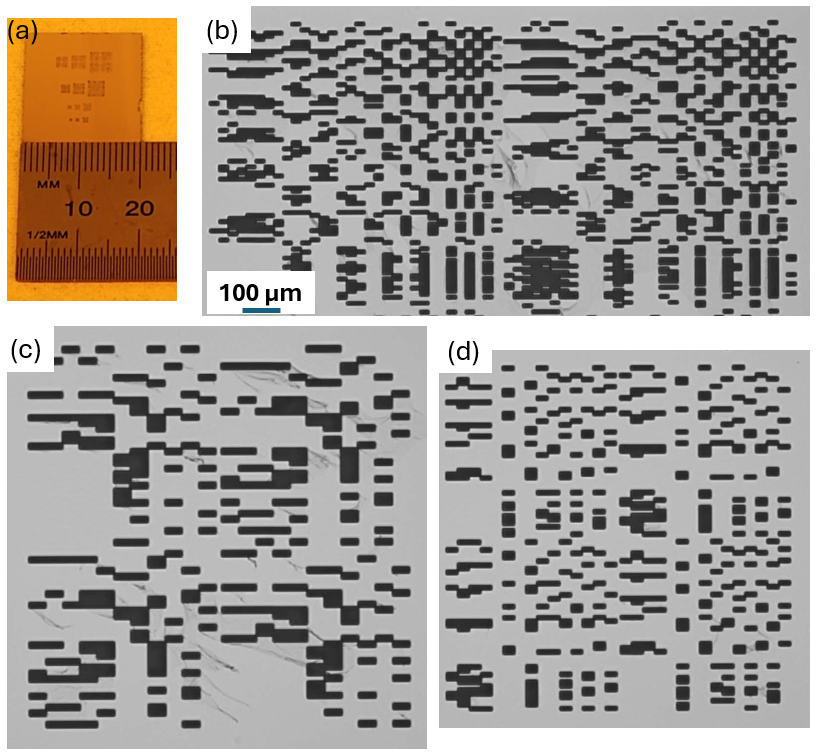}\\
\caption{(a) One of the fabricated 2cm $\times$ 2cm substrates containing 12 binary Huffman-like masks. Optical images of (b) a $PN$ pair of the fabricated $32\times32$ binary Huffman-like mask with 8 $\mu$m resolution, (c) a fabricated $11\times11$ binary Huffman-like mask with 15 $\mu$m resolution, and (d) a fabricated $15\times15$ binary Huffman-like mask with 10 $\mu$m resolution. The scale bar shown in b) also applies for the (c) and (d) images. The streak artefact in the image are residue from photoresist, which are transparent under x-rays.}
 \label{fig:Ternary images}
\end{figure}

Binary masks were fabricated on 2cm $\times$ 2cm $\textrm{SiO}_2$ wafers, which were coated with a 5 $\mu$m thick Ta layer. 5 $\mu$m was, approximately, the maximum achievable film thickness using the sputtering technique, and Ta had relatively low x-ray transmissions (among the available materials) over the selected photon energies. More information is provided in the Supplementary Material. An example fabricated wafer is depicted in Fig.~\ref{fig:Ternary images}(a). It includes 12 masks with four sizes (i.e. $11\times11$, $15\times15$, $32\times32$, and $43\times43$) and three different resolutions (i.e. 8 $\mu$m, 10 $\mu$m, and 15 $\mu$m array subpixel sizes).

Optical images of the $[P,N/N,P]$ masks for the $32 \times 32$ arrays with 8 $\mu$m pixel resolution, the $11 \times 11$ mask with 15 $\mu$m resolution, and the $15 \times 15$ mask with 10 $\mu$m resolution, are illustrated in Fig.~\ref{fig:Ternary images}(b)-(d). Note that the streak artefacts in the optical images are residue from the photoresist, which are near-transparent under x-rays and do not affect the functionality of the masks.

The fabrication process for the quaternary masks was more complex, although the same very large scale integration (VLSI) techniques were utilised. Quaternary Huffman-like masks require four levels $\{ 0, 1, 2, 3 \}$ with each level transmitting x-rays in steps of uniformly increasing intensity. Pixels at each level of the quaternary mask can be fabricated with a specific uniform thickness of Ta to provide the required level of x-ray transmission. Given the maximum achievable thickness of approximately 5 $\mu$m using the sputtering technique (as discussed in the Supplementary Materials), we can estimate the minimum x-ray transmission through our mask as approximately 17.5$\%$ at 12.4 keV; this is the transmission through level 0. 12.4 keV is just above the L absorption edges of Ta providing maximum attenuation of Ta while still allowing transmission through silica. Requiring x-ray transmission, $T$, the thickness, $t$, of each level can be measured as $t = -ln(T)/\mu$, where $\mu$ is the linear attenuation coefficient (0.3605 $\mu\mathrm{m}^{-1}$ for Ta at 12.4 keV \cite{bergerXcom2010}). The x-ray transmission and Ta thickness for each level is shown in Table \ref{tab:Tr for each level}.

\begin{table}
    \centering
    \begin{tabular}{ccc}
        \toprule
       Layers/Levels  & X-ray Transmission at 12.4 keV & Ta Thickness\\
        \midrule
        0 & 17.5$\%$ & 4.8$\mu$m\\
        1 & 45$\%$ & 2.2$\mu$m\\
        2 & 72.5$\%$ & 0.88$\mu$m\\
        3 & 100$\%$ & 0\\
        \bottomrule
    \end{tabular}
    \caption{X-ray transmissions and Ta thicknesses for each level of the quaternary Huffman-like masks}
    \label{tab:Tr for each level}
\end{table}

\begin{figure}[h!]
    \centering
    \includegraphics[width=0.9\linewidth] {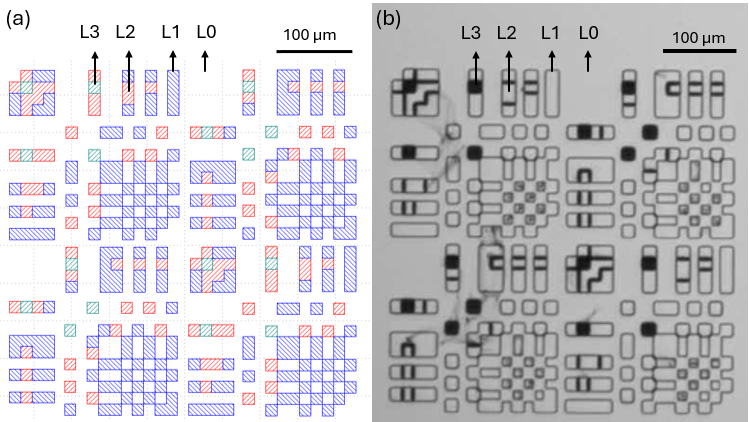}\\
\caption{(a) design and (b) optical image of the fabricated $15\times15$ quaternary Huffman-like mask. L0, L1, L2, and L3 represent level 0, 1, 2, and 3 of the mask respectively.}
 \label{fig:Fabricated Huffman}
\end{figure}

 Levels 1, 2 and 3 were fabricated through three lithography steps (one for each level) followed by three etching steps. Different etching times were required to achieve different thicknesses (see Supplementary Material for more detail). We fabricated 15 quaternary masks on a 4-inch $\textrm{SiO}_2$ wafer. The masks were fabricated with five array sizes ($11 \times 11$, $15 \times 15$, $32 \times 32$, $43 \times 43$, and $86 \times 86$) and three resolutions (10 $\mu$m, 15 $\mu$m, and 20 $\mu$m pixel sizes).  An optical image of the fabricated $15 \times 15$ quaternary mask is depicted in Fig.~\ref{fig:Fabricated Huffman}(b). The four Ta levels are indicated in the image. Note that level 0 (L0) is the substrate, which was coated with approximately 5 $\mu$m Ta.

\section{Validation of Mask Fabrication}
\label{sec: Mask Validation}

Before employing the masks in diffuse scanning probe experiments, the masks were examined at the Micro-Computed Tomography (MCT) beamline at the Australian Synchrotron. We validated the quality of mask fabrication by imaging the masks using a 2D pixelated x-ray detector using 20 keV and 12.4 keV x-ray illumination. Based on the transmitted intensity images, we assessed the faithfulness of the mask structure and uniformity of transmission levels compared with the ideal array designs.

\subsubsection{Binary Masks} \label{sec: Binary Huffman-like Masks}

\begin{figure}
    \centering
    \begin{minipage}{0.5\linewidth}
    \centering
    \scriptsize{(a)}\\
    \includegraphics[width=0.9\linewidth]{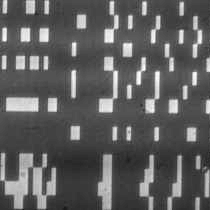}
    \end{minipage}%
    \begin{minipage}{0.5\linewidth}
    \centering
    \scriptsize{(b)}\\
    \includegraphics[width=0.9\linewidth]{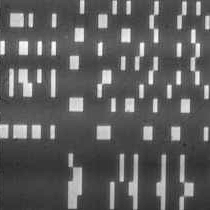}
    \end{minipage}\\
    \begin{minipage}{0.5\linewidth}
    \scriptsize{(c)}\\
    \includegraphics[width=0.9\linewidth]{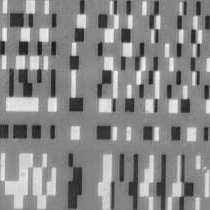}
    \end{minipage}%
    \begin{minipage}{0.5\linewidth}
    \centering
    \scriptsize{(d)}\\
    \includegraphics[width=\linewidth]{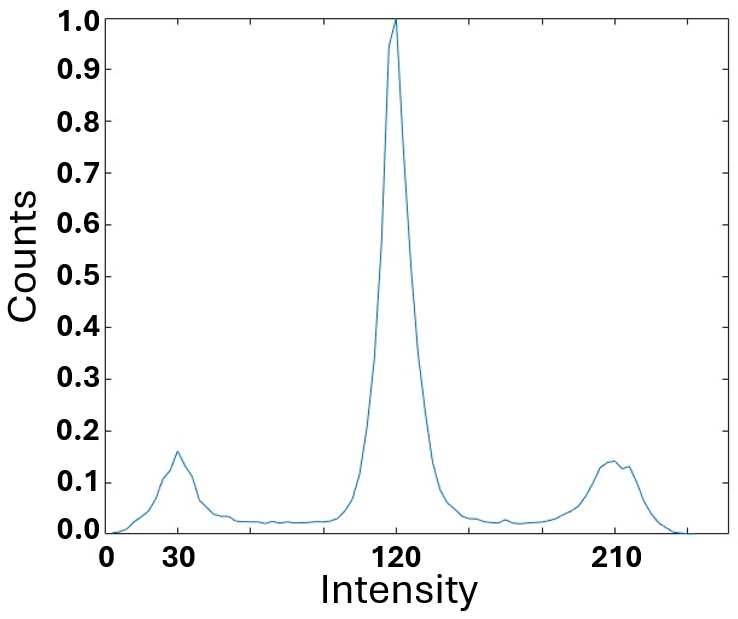}
    \end{minipage}%
    \caption{Detector images of the (a) positive ($P$) and (b) negative ($N$) $15 \times 15$ pixel Huffman-like mask generated using $3 \times 3$ binary pixels of pitch 15$\mu$m. (c) the signed mask, re-generated as $P-N$. (d) shows the histogram of image (c). The three peaks show the distribution of the measured x-ray illumination intensities that passed through the $-1, 0, +1$ sections of the fabricated mask, respectively. The vertical axis in (d) is counts, the horizontal axis displays the relative signal intensity.} 
    \label{fig:binary_mask_expt_example}
\end{figure}

An example of the binary mask transmission pattern for the $15 \times 15$ pixel $P$ and $N$ regions with 15 $\mu$m subpixel pitch is shown in Figs. \ref{fig:binary_mask_expt_example}(a) and \ref{fig:binary_mask_expt_example}(b). The ``reassembled'' Huffman-like array, calculated as $P-N$, is presented in Fig. \ref{fig:binary_mask_expt_example}(c). The histogram of measured x-ray intensities transmitted in this reassembled array are presented in Fig.~\ref{fig:binary_mask_expt_example}(d). Here the x-ray beam energy was $20$ keV with a 3\% bandpass. 

\subsubsection{Quaternary Masks}\label{sec:Quaternary Huffman-like Masks}

Images were acquired of the transmission of x-rays through the $[P,N / N,P]$ mask arrangement for the $11 \times 11$ quaternary mask with 10 $\mu$m and 20 $\mu$m pixels, for the $15 \times 15$ quaternary mask with 10 $\mu$m and 20 $\mu$m pixels and for the $32 \times 32$ mask with 20 $\mu$m pixels. Here the x-ray beam energy was $12.4$ keV with 1\% bandpass to reduce variation in the energy-dependent transmission.

\begin{figure}
    \centering
    \begin{minipage}{0.33\linewidth}
    \centering
    \scriptsize{(a)}\\
    \includegraphics[width=0.95\linewidth]{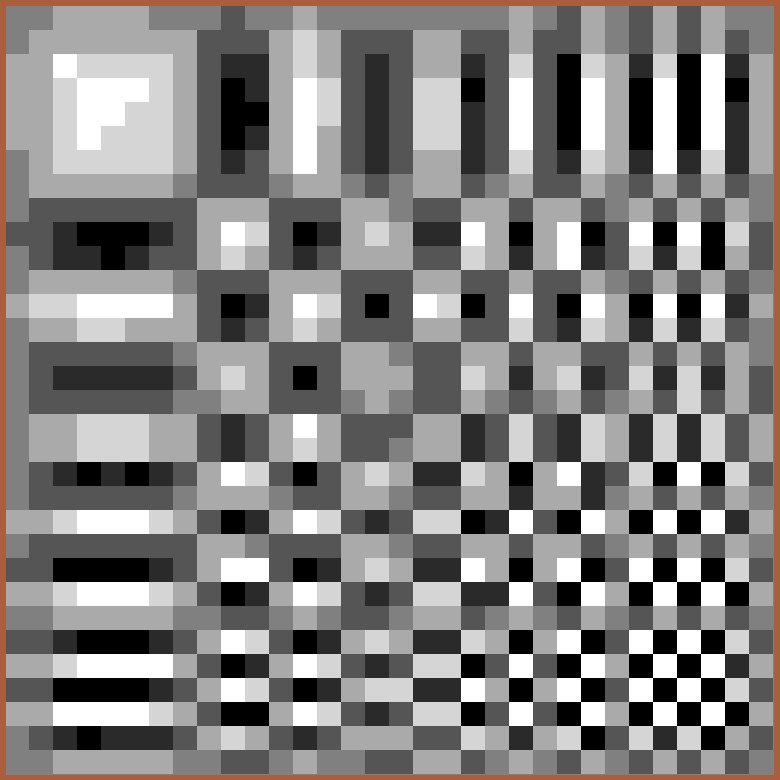}
    \end{minipage}%
    \begin{minipage}{0.33\linewidth}
    \centering
    \scriptsize{(b)}\\
    \includegraphics[width=0.95\linewidth]{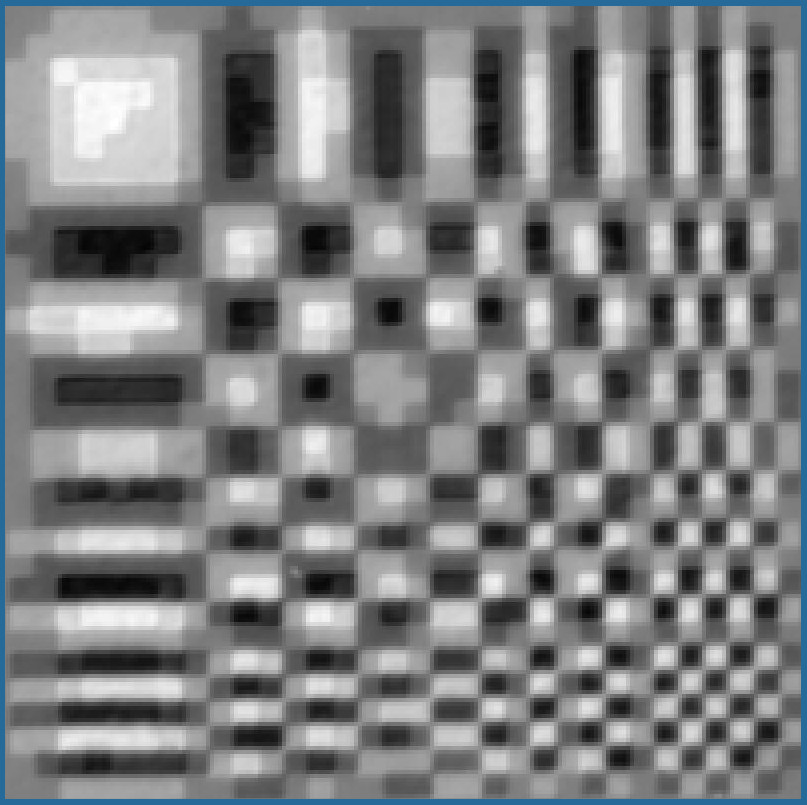}
    \end{minipage}%
    \begin{minipage}{0.34\linewidth}
    \centering
    \scriptsize{(c)}\\
    \includegraphics[width=\linewidth]{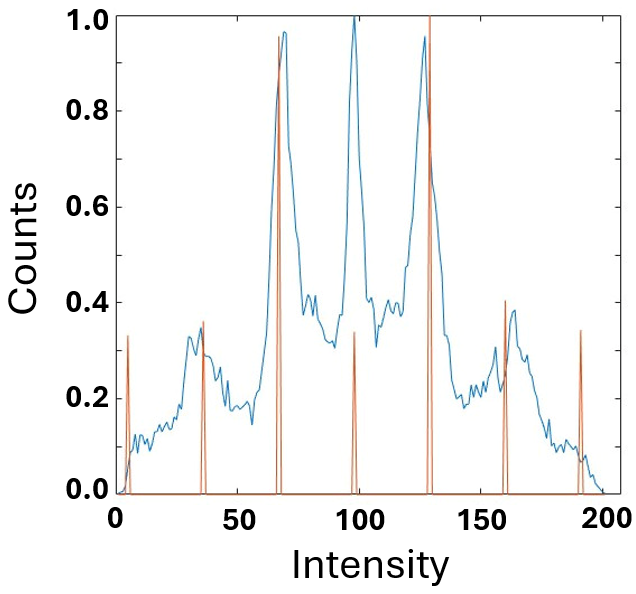}
    \end{minipage}
    \caption{(a) Image of an ideal $32\times 32$ $\pm 3$ gray-level Huffman-like mask, (b) Image of x-ray transmission through the fabricated mask with 20 $\mu$m pixel size, presented as $P - N$. (c) The histogram of graylevels in the ideal mask, image (a) (shown in orange), registered with the histogram of (b) the intensity of the x-rays transmitted through the fabricated mask (shown in blue). Both image histograms show five closely-matched peaks, representing the relative x-ray transmission intensities for mask levels $-2, -1, 0, +1, +2$ . The vertical scale is counts, the horizontal axis is relative intensity.}
    \label{fig:quaternary_expt_example}
\end{figure}

As an example, the $32 \times 32$ Huffman-like array is presented in Fig.~\ref{fig:quaternary_expt_example}(a). The reassembled array generated from the $P$ and $N$ x-ray images of the $32\times 32$ quaternary mask with 20 $\mu$m pixel size is presented in Fig.~\ref{fig:quaternary_expt_example}(b) for comparison. The mask appears to produce a faithful representation of the ideal array. The histogram of this reassembled Huffman-like array is presented in Fig.~\ref{fig:quaternary_expt_example}(c). That histogram closely follows the spacing and relative intensities of the histogram of the ideal array, shown as the seven narrow peaks scaled and overlaid on the plot of measured x-ray intensities. The higher frequency observed around the central ``zero'' intensity peak results from the entries generated along the edges of pixels where the measured images of the fabricated $P$ and $N$ meet and have their intensities subtracted to form the $P-N$ image (hence creating extra zero pixels).

\begin{figure}
    \centering
    \begin{minipage}{0.5\linewidth}
    \scriptsize{(a)}\\
    \includegraphics[height=0.95\linewidth]{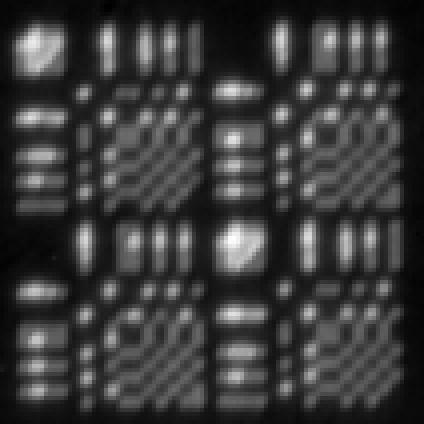}
    \end{minipage}%
    \begin{minipage}{0.5\linewidth}
    \centering
    \scriptsize{(b)}\\
    \includegraphics[height=0.95\linewidth]{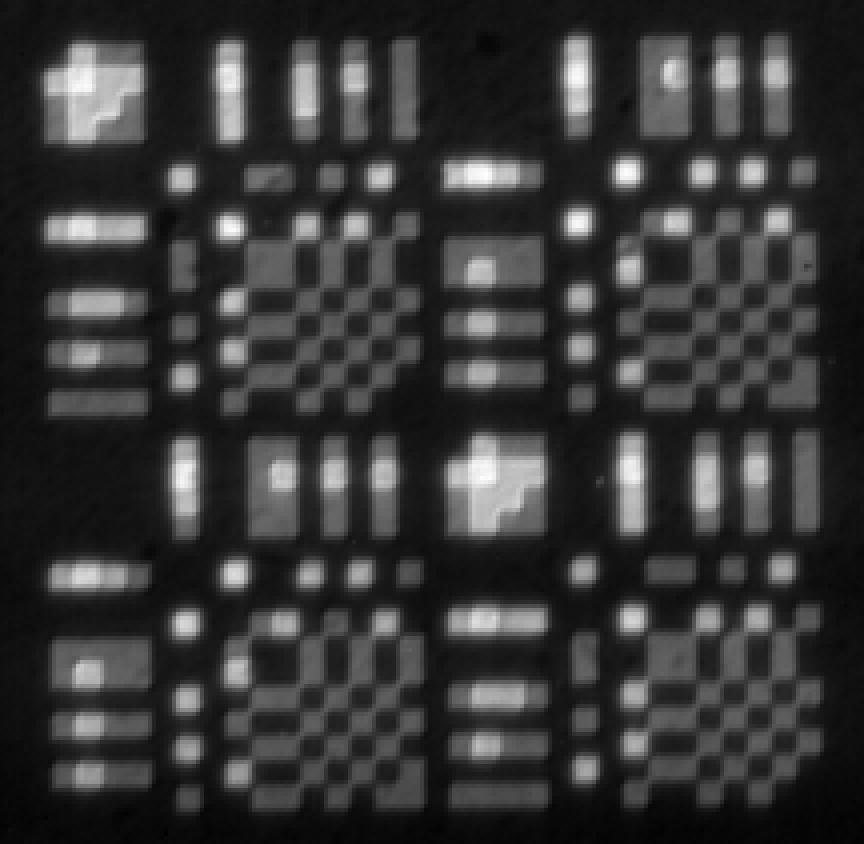}
    \end{minipage}
    \caption{
    (a) Image of x-ray transmission through the $[P,N / N,P]$ fabricated $15\times15$ gray-level masks with (a) $10$ $\mu$m pixels and (b) $20$ $\mu$m pixels.}
    \label{fig:quaternary_PNNP_expt}
\end{figure}

\begin{figure}
    \centering
    \begin{minipage}{0.33\linewidth}
    \scriptsize{(a)}\\
    \includegraphics[height=0.95\linewidth]{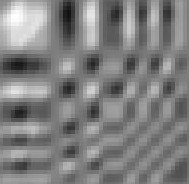}
    \end{minipage}%
    \begin{minipage}{0.33\linewidth}
    \centering
    \scriptsize{(b)}\\
    \includegraphics[height=0.95\linewidth]{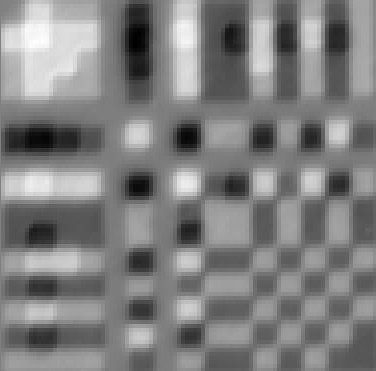}
    \end{minipage}%
    \begin{minipage}{0.33\linewidth}
    \scriptsize{(c)}\\
    \includegraphics[height=0.95\linewidth]{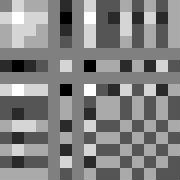}
    \end{minipage}
    \caption{
    The $15\times15$ gray-level masks formed from Fig.~\ref{fig:quaternary_PNNP_expt} as $P-N$ with (a) $10$ $\mu$m  pixels, (b) $20$ $\mu$m  pixels. (c) the ideal mask for comparison.}
    \label{fig:quaternary_expt}
\end{figure}

Figure \ref{fig:quaternary_PNNP_expt} shows two example ``raw'' x-ray transmission images of the quaternary masks for a $15 \times 15$ array at two different pixel sizes ($10$ $\mu$m and $20$ $\mu$m). The mask image data was acquired using a long exposure (or was summed over multiple repeated exposures) to reduce noise. The reassembled Huffman-like arrays using these masks are presented in Figs. \ref{fig:quaternary_expt}(a) and \ref{fig:quaternary_expt}(b) to be compared with the ideal array in Fig.~\ref{fig:quaternary_expt}(c). Strong agreement between each original and fabricated mask is observed. Ta thickness here was changed by three successive depositions and etchings, resulting in some small variations around the pixel edges. These variations are much less evident for the 20 $\mu$m pixel fabrication than for the 10 $\mu$m pixel fabrication.

%


\section{Experimental scanning probe method}
\label{sec:Experimental Method}

\subsection{Experiment set-up}
\label{sec:exp_setup}

The masks were employed to pattern diffuse scanning probes at the MCT beamline at the Australian Synchrotron. A monochromatic x-ray beam with energy of 20 keV was first used to analyse the performance of the binary masks. Note that the binary masks can be used across a range of photon energies, since they contain only opaque and transparent parts. In contrast, the quaternary masks require each level to transmit a specific percentage of the incident beam. The thicknesses of Ta to give each transmission fraction were optimised for a single photon energy (12.4 keV in this case).

\begin{figure}[h!]
    \centering
    \includegraphics[width=\linewidth]{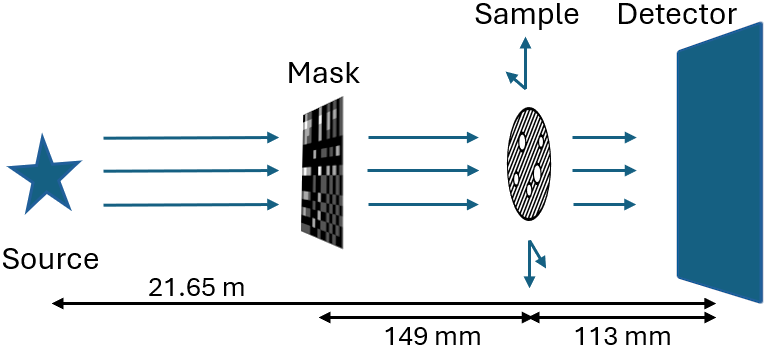}\\
\caption{Experimental geometry on the MCT beamline at the Australian Synchrotron. The mask, sample translation and detector planes define the coordinates $(x,y)$, while the beam propagates in the $+z$ direction.}
 \label{fig:Exp geometry}
\end{figure}

A schematic of the experimental setup is shown in Fig.~\ref{fig:Exp geometry}. The mask was placed approximately 262 mm from the detector, and the sample was located approximately 149 mm in front of the mask. The sample was mounted on a moving stage to allow the test object to be transversely scanned over the selected mask. A 2D pixelated detector was utilised throughout the entire process. The resulting images were then processed digitally into 4 individual sensors (or buckets), one for each quadrant of the $[P,N/N,P]$ arrays. As each scan proceeded, these bucket values were collected into bucket arrays for image reconstruction, as described in Sec.~\ref{sec:exp_procedure}.

\subsection{Test Objects}
\label{sec:Test Objects}

The performance of the Huffman-like probes was examined with a range of test objects specifically designed and fabricated for this purpose. We started with simple pinholes and progressed to more complex objects. Scanning a pinhole test object is of interest as the resulting bucket image should reproduce an exact image of the discretely-shaped x-ray beam intensity. This image is obtained with very weak illumination, as the entire bucket receives information from just one mask pixel at each scan point. The pinhole imaging results test the resilience of broad Huffman-like array imaging under very low signal-to-noise conditions. 

\begin{figure}[h!]
    \centering
    \includegraphics[width=0.8\linewidth]{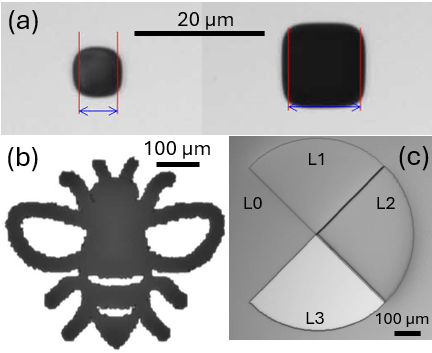}\\
\caption{Optical images of some of the fabricated test objects: (a) 5 $\mu$m and 10 $\mu$m pinholes, (b) a binary object (bee image) and (c) disc with four quadrants, each quadrant having a different uniform thickness ($ L3 < L2 < L1 < L0$).}
 \label{fig:Test objects}
\end{figure}

Four pinholes, with diameters of 5 $\mu$m, 10 $\mu$m, 15 $\mu$m, and 20 $\mu$m, were fabricated to be imaged using different mask sizes. Optical images of two fabricated pinholes are illustrated in Fig.~\ref{fig:Test objects}(a). The material and fabrication process were the same as those used for the binary Huffman-like mask explained in Sec. \ref{sec:mask_fab}. Utilising the same procedure, multiple binary objects of different sizes were also fabricated. An optical image of a binary test object is shown in Fig.~\ref{fig:Test objects}(b). We also fabricated gray-level objects using the four-level fabrication method described in Sec. \ref{sec:mask_fab}. Figure.~\ref{fig:Test objects}(c) shows an example of a gray-level test object, where each quadrant of the circle has a unique, uniform thickness of Ta.

\subsection{Experimental procedure}
\label{sec:exp_procedure}

The object needs to be over-scanned on each edge by at least the full size of the mask to obtain valid bucket signals able to accurately deconvolve up to the object edges. Thus the acquisition time for 2D images and the size of the required bucket data scales with both the object size, desired spatial resolution, and the chosen array size of the mask. 

We first scanned different pinholes (one at a time) over a selection of binary masks. As described above, the number of images collected for each scan depended on the array size of the masks. For instance, $31 \times 31$ images were collected for a $15 \times 15$ array mask to cover the entire $[P,N/N,P]$ mask. The exposure time was 0.03 seconds for each position in these scans. In addition to the scans, we collected 10 flat-field (FF) and 10 dark-field (DF) images, as well as 10 mask-field (MF) images i.e. images of the mask only. Utilising the binary masks, we also scanned other test objects including the bee and the gray-level circle, as shown in Fig.~\ref{fig:Test objects}(b) and (c) respectively.

After scanning multiple objects with binary masks using 20 keV x rays, we changed the photon energy to 12.4 keV and repeated similar experiments using the quaternary masks. The exposure time was increased to 0.2 seconds to improve the signal-to-noise ratio (SNR). We first scanned a pinhole over several $11 \times 11$ quaternary masks and then progressed to $15 \times 15$ and $32 \times 32$ masks. In addition to the pinholes, a few binary and gray-level objects were scanned using $11 \times 11$ and $15 \times 15$ quaternary masks. Again, 10 images of the FF, DF, and MF were also collected for each scan.

\begin{figure}[h!]
    \centering
    \begin{minipage}{0.79\linewidth}
    \centering
    \scriptsize{(a)}\\
    \includegraphics[width=\linewidth]{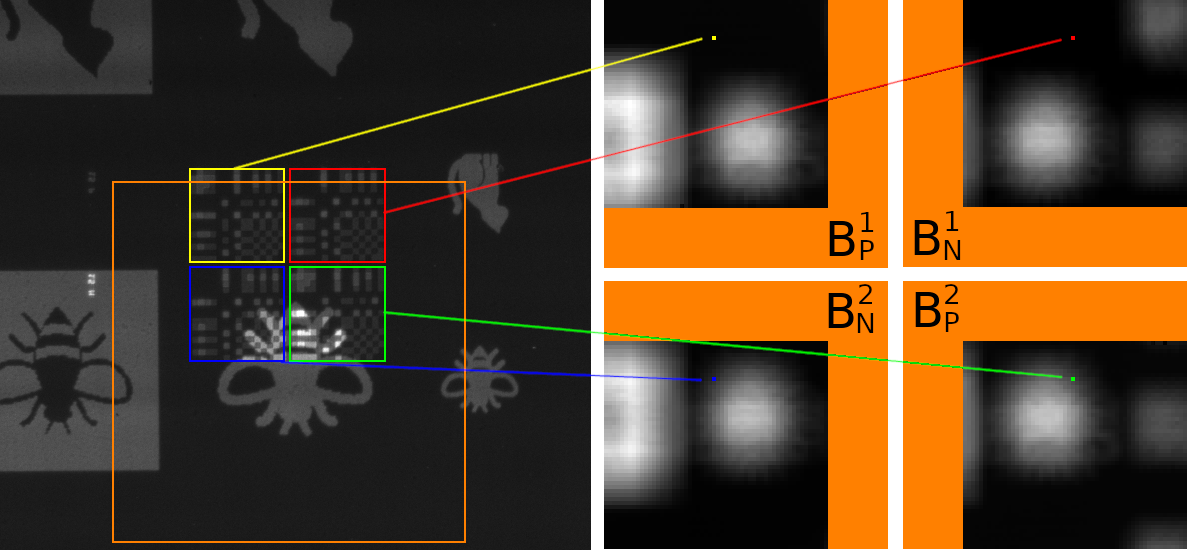}
    \end{minipage}%
    \begin{minipage}{0.21\linewidth}
    \centering
    \scriptsize{(b)}\\
    \includegraphics[width=0.95\linewidth]{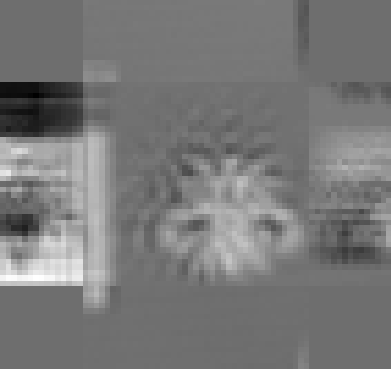}
    \end{minipage}
\caption{An example of the process to generate bucket images. A raw experimental radiograph is shown on the left in (a) with the four mask quadrants identified using coloured squares. The centre of the mask region is scanned around in the orange square. At each position integrated pixel values in each quadrant are added to the respective pixels of the bucket images, $\{B^1_P, B^1_N, B^2_N, B^2_P\}$, for each quadrant on the right. Note the $P$ and $N$ masks individually produce bucket values that form a low-pass filtered image of the object being scanned. The positive, $\{B^1_P, B^2_P\}$, and negative, $\{B^1_N, B^2_N\}$, bucket image pairs are combined and subtracted to form the diffuse-probe scanned image shown in (b). This image is deconvolved to produce the result in Fig.~\ref{fig:binary_results}(a-iii)}
 \label{fig:bucket_image_creation}
\end{figure}

Once the sets of 2D pixelated images were collected for each scan, a series of postprocessing analysis was performed to create four bucket images for each scan:
\begin{enumerate}
    \item Average the 10 FF and 10 DF images.
    \item Subtract the averaged DF image from both the FF and scanned images.
    \item Normalise the scanned images by dividing by the FF image.
    \item Identify the coordinates of each mask quadrant (see coloured squares in an example radiograph in Fig.~\ref{fig:bucket_image_creation}(a)).
    \item Integrate the measured intensities over each quadrant into separate bucket images $B_N$ and $B_P$ (depicted for an example raw experimental radiograph in Fig.~\ref{fig:bucket_image_creation}(a)).
    \item Combine the aligned and normalised $P$ and $N$ bucket images as $B_P-B_N$ to give the result of scanning the object with a Huffman-like array (depicted for an example in Fig.~\ref{fig:bucket_image_creation}(b)).
    \item Deconvolve the combined bucket image using the ideal Huffman-like array to reconstruct an image of the scanned sample.
\end{enumerate}

\section{Experimental results: reconstructed x-ray images}
\label{sec:Experimental results}

Here we present the results obtained using diffuse x-ray scanning probes shaped by our masks that are based on Huffman-like arrays. The test object images are reconstructed from the scanned data (or {\it bucket signals}) obtained using both binary and quaternary fabricated masks of different sizes and spatial resolutions. The scans are performed by raster scanning the objects across the stationary patterned illumination probe. Results for the $15 \times 15$ binary mask and $32 \times 32$ quaternary mask are shown in the following sections. In both cases, the first object is a pinhole aperture with dimensions similar to that of the mask pixels. Following that, simple binary objects were scanned, and finally multilevel (gray) objects were imaged.

\subsubsection{Pinhole Images: Binary Mask}
\label{sec:Binary Huffman-like Mask Pinhole Images}

\begin{figure}
    \centering
    \begin{minipage}{0.58\linewidth}
    \centering
    \scriptsize{(a)}\\
    \includegraphics[width=0.95\linewidth]{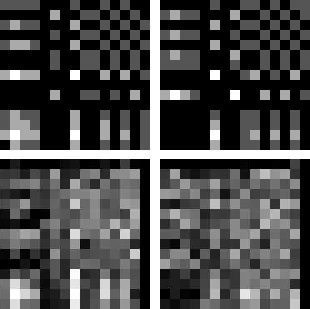}
    \end{minipage}%
    \begin{minipage}{0.42\linewidth}
    \centering
    \scriptsize{(b)}\\
    \includegraphics[width=0.95\linewidth]{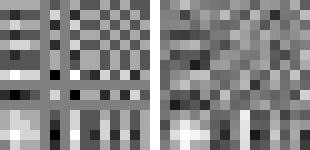}\\
    \scriptsize{(c)}\\
    \includegraphics[width=\linewidth]{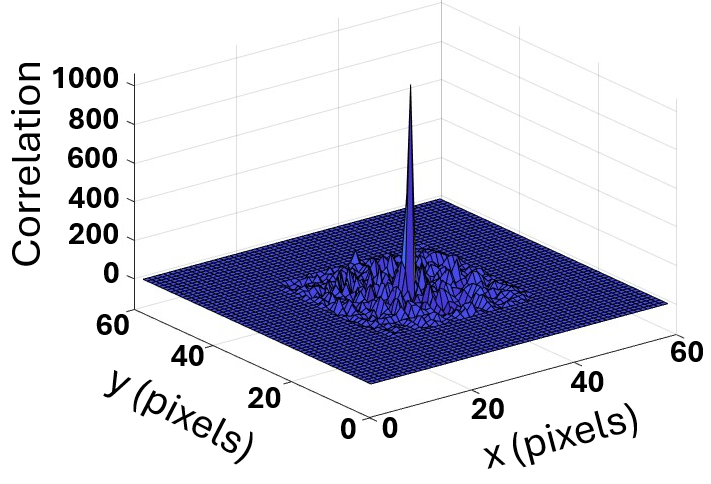}
    \end{minipage}
    \caption{ Pinhole images of the binary $15 \times 15$ mask. (a) Top row: original $P$ and $N$ integer arrays, black = 0, white = 3. Bottom row: x-ray bucket images obtained for the $P$ and $N$ mask. (b) Left: original Huffman-like array, black = -3, white = +3. Right: reconstructed x-ray pinhole image of the $15 \times 15$ Huffman-like mask. (c) A surface plot of the cross-correlation of the original integer array (left image of (b)) with the reconstructed x-ray-image of the mask (right image of (b)). The correlation is delta-like. The vertical scale has been normalised to set the maximum of the central peak to have value $100$.}
    \label{fig:ternaryPinholeResults}
\end{figure}

The $15 \times 15$ binary mask, in $[P,N/N,P]$ form, has $P$ and $N$ masks with $45 \times 45$ subpixels (due to the ${3\times3}$ subpixel area-weighting used to accommodate the gray levels $[0, 1, 2, 3]$). The binary mask, fabricated with 8 $\mu$m subpixels, was used to illuminate a 20 $\mu$m pinhole. This pinhole almost covers the physical 24 $\mu$m width of the fullsize mask pixel, i.e., ${3\times3}$ subpixels. The pinhole was 2D raster scanned, using 24 $\mu$m steps, to produce a $15 \times 15$ bucket image. Figure~\ref{fig:ternaryPinholeResults}(a) shows the ideal $P$ and $N$ integer arrays and, below, the mean of the two $P$ and two $N$ x-ray bucket images. On the left of Fig.~\ref{fig:ternaryPinholeResults}(b) is the ideal Huffman-like ${\pm3}$ valued array for comparison with the the reassembled x-ray bucket image, i.e., $P-N$, show on the right. The x-ray pinhole mask image clearly resembles the original Huffman-like array. The surface plot in Fig.~\ref{fig:ternaryPinholeResults}(c) shows the reassembled Huffman-like image cross-correlated with the ideal array.

The binary masks with 8 $\mu$m subpixels were fabricated with an 8 $\mu$m gap separating the $P$ and $N$ boundaries of the $[P,N/N,P]$ layout. The 24 $\mu$m steps used to scan the pinhole were thus spatially misaligned by a third of a pixel after stepping across the $P$ and $N$ boundary gaps. The raster scan axis was also rotated by about a degree relative to the mask axis, meaning the scan and mask locations had some shear. The end-points of the raster scan were terminated nearly a pixel short of the mask boundaries, as is evident in Fig.~\ref{fig:ternaryPinholeResults}. Despite the imaging misalignment and the low signal-to-noise ratio, the correlation of this diffuse Huffman-like scanned image with the original array proved to be delta-like, reconstructing the pinhole faithfully (see Fig.~\ref{fig:ternaryPinholeResults}(c)).

\subsubsection{Pinhole Images: Quaternary Mask}
\label{sec:Quaternary Huffman-like Mask Pinhole Images}

\begin{figure}
    \centering
    \begin{minipage}{0.25\linewidth}
    \centering
    \scriptsize{(a)}\\
    \includegraphics[width=0.95\linewidth]{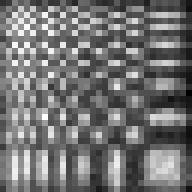}
    \end{minipage}%
    \begin{minipage}{0.25\linewidth}
    \centering
    \scriptsize{(b)}\\
    \includegraphics[width=0.95\linewidth]{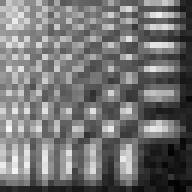}
    \end{minipage}%
    \begin{minipage}{0.25\linewidth}
    \centering
    \scriptsize{(c)}\\
    \includegraphics[width=0.95\linewidth]{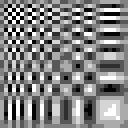}
    \end{minipage}%
    \begin{minipage}{0.25\linewidth}
    \centering
    \scriptsize{(d)}\\
    \includegraphics[width=0.95\linewidth]{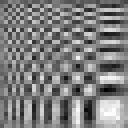}
    \end{minipage}
    \caption{ Pinhole images of the $32 \times 32$ quaternary mask. (a) mean x-ray bucket image obtained for the $P$ mask. (b) mean bucket image for the $N$ mask. (c) original Huffman-like array, black = -3, white = +3. (d) reconstructed x-ray pinhole image of the $32 \times 32$ Huffman-like mask.
    }
    \label{fig:greyscalePinholeResults}
\end{figure}

The $32 \times 32$ quaternary mask was fabricated with 20 $\mu$m subpixels. This mask was used to illuminate a 20 $\mu$m pinhole with a beam of 12.4 keV x-rays. The pinhole was 2D raster scanned in 20 micron steps to produce $32 \times 32$ $P$ and $N$ bucket images. The mean of the two resulting $P$ and mean of the two $N$ x-ray bucket image measurements are shown in Fig.~\ref{fig:greyscalePinholeResults}(a) and (b) respectively. These bucket images were reassembled, as $P-N$, to recover the Huffman-like scanning probe image, as shown in Fig.~\ref{fig:greyscalePinholeResults}(d). The x-ray Huffman-like probe image very closely resembles the ideal Huffman-like array depicted in Fig.~\ref{fig:greyscalePinholeResults}(c).

The cross-correlation of the $32 \times 32$ Huffman-like x-ray probe image with the ideal array was delta-like, thereby reconstructing the pinhole well. The values in the area around the peak are given in Table \ref{tab:5x5zoom}. This result confirms that the broad x-ray probe (with a footprint spread over an area of 1024 pixels) is indeed able to be sharply focused to a pixel-like point when deconvolved.

\begin{table}[h!]
\begin{center}
\begin{tabular}{| c | c | c | c | c |}
\hline
-0.4	&	-1.6	&	-4.9	&	-2.0	&	0.8	\\
\hline
0.0	&	6.6	&	5.9	&	0.8	&	0.2	\\
\hline
-3.1	&	22.6	&	{\bf 100.0}	&	19.0	&	-0.4	\\
\hline
-0.4	&	13.2	&	23.8	&	0.2	&	0.1	\\
\hline
2.0	&	4.5	&	1.1	&	1.6	&	0.6	\\
\hline
\end{tabular}
\end{center}
\caption{
Values
for the $5\times5$ area around the central peak of the cross-correlation of the $32 \times 32$ mask x-ray pinhole image with the ideal Huffman-like array.
Values are given as the percentage of the central peak shown in bold.}
\label{tab:5x5zoom}
\end{table}

\subsubsection{Binary and Gray-level Object Images: Binary Mask}
\label{sec:Binary and Grey-level Object Images}

\begin{figure}
    \centering
    \begin{minipage}{0.33\linewidth}
    \centering
    \scriptsize{(a-i)}\\
    \includegraphics[width=0.9\linewidth]{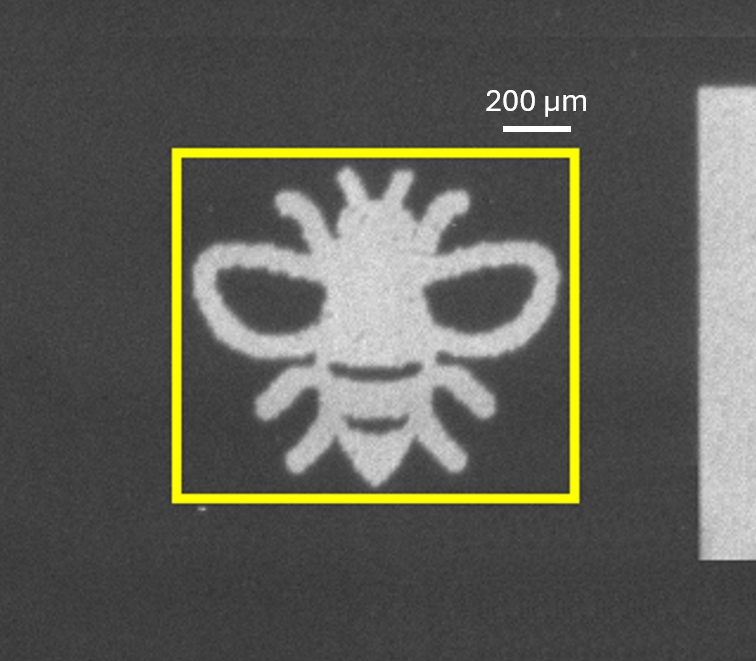}
    \end{minipage}%
    \begin{minipage}{0.33\linewidth}
    \centering
    \scriptsize{(a-ii)}\\
    \includegraphics[width=0.9\linewidth]{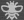}
    \end{minipage}%
    \begin{minipage}{0.33\linewidth}
    \centering
    \scriptsize{(a-iii)}\\
    \includegraphics[width=0.9\linewidth]{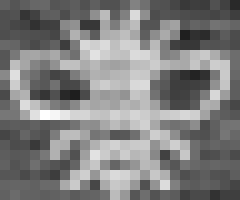}
    \end{minipage}\\
    \begin{minipage}{0.33\linewidth}
    \scriptsize{(b-i)}\\
    \includegraphics[width=0.9\linewidth]{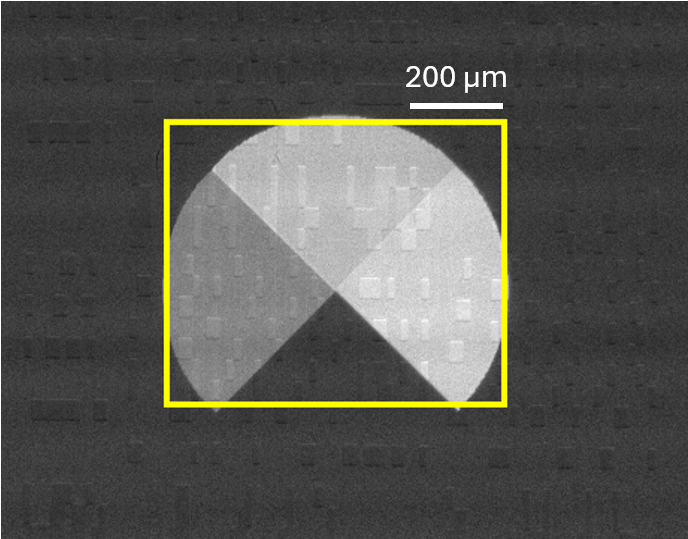}
    \end{minipage}%
    \begin{minipage}{0.33\linewidth}
    \centering
    \scriptsize{(b-ii)}\\
    \includegraphics[width=0.9\linewidth]{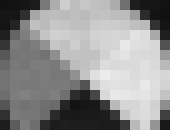}
    \end{minipage}%
    \begin{minipage}{0.33\linewidth}
    \scriptsize{(b-iii)}\\
    \includegraphics[width=0.9\linewidth]{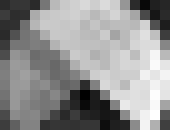}
    \end{minipage}
    \caption{(a) Binary image of a bee, and (b) multiple gray-level circle with quadrants of different intensities. (iii) Recovered images compared with (ii) the expected images  using a quaternary $15 \times 15$ Huffman like array created using $3 \times 3$ binary subpixels with pitch 15 $\mu$m.}
    \label{fig:binary_results}
\end{figure}

Figure~\ref{fig:binary_results}(a-i) shows, in the highlighted region, an x-ray radiograph of a binary object (a ``bee'') to give context of the surrounding region that is also included to some degree in the scan. Figure~\ref{fig:binary_results}(a-ii) shows this region rescaled and binned to the same field-of-view and pixel size as the reconstructed image in Fig.~\ref{fig:binary_results}(a-iii). This last image is the Huffman-like x-ray probe image obtained from the $15 \times 15$ binary mask (with 15 $\mu$m pixels) scanned over this object, after being deconvolved by the ideal Huffman-like array. The radiograph in Fig.~\ref{fig:binary_results}(a-ii) and reconstructed diffuse scanning-probe image in Fig.~\ref{fig:binary_results}(a-iii) show close agreement. 

Figure~\ref{fig:binary_results}(b-i) shows an x-ray radiograph of a circular disc, printed with subquadrants of different but uniform Ta thickness to provide a (gray-level) range of transmitted intensities. Figure~\ref{fig:binary_results}(b-ii) shows this region rescaled and binned to the same field-of-view and pixel size as the reconstructed image in Fig.~\ref{fig:binary_results}(b-iii). Figure~\ref{fig:binary_results}(b-iii) was obtained after deconvolving the $P-N$ Huffman-like x-ray probe image obtained by scanning the $15 \times 15$ binary mask (with 15 $\mu$m subpixels) over this ``gray'' disc. As for the binary mask results, the radiograph in Fig.~\ref{fig:binary_results}(b-ii) and reconstructed diffuse scanning-probe image in Fig.~\ref{fig:binary_results}(b-iii) show close agreement. 

\subsubsection{Binary and Gray-level Object images: Quaternary Mask}
\label{sec:Quaternary Mask Binary and Grey-level Object images}

\begin{figure}
    \centering
    \begin{minipage}{0.33\linewidth}
    \centering
    \scriptsize{(a-i)}\\
    \includegraphics[width=0.9\linewidth]{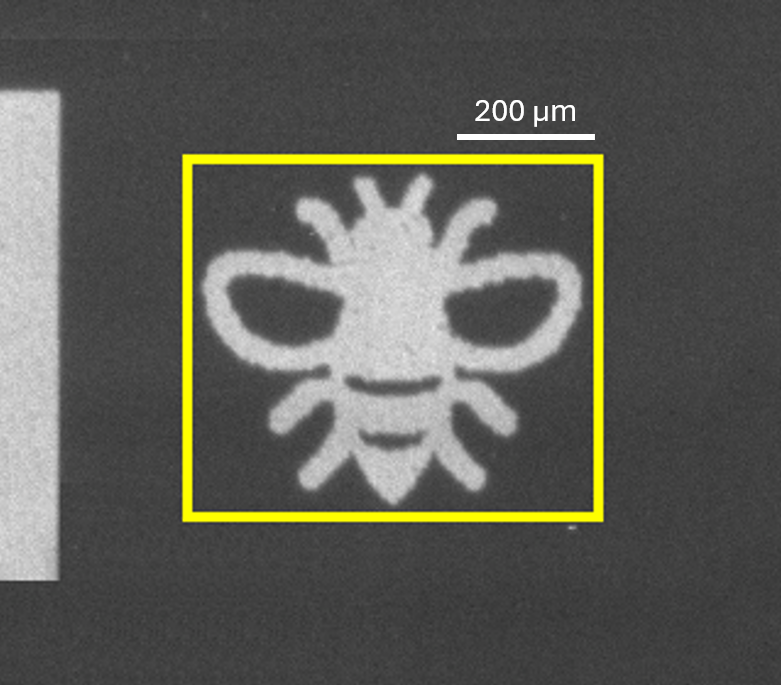}
    \end{minipage}%
    \begin{minipage}{0.33\linewidth}
    \centering
    \scriptsize{(a-ii)}\\
    \includegraphics[width=0.9\linewidth]{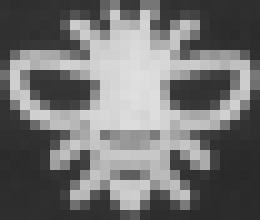}
    \end{minipage}%
    \begin{minipage}{0.33\linewidth}
    \centering
    \scriptsize{(a-iii)}\\
    \includegraphics[width=0.9\linewidth]{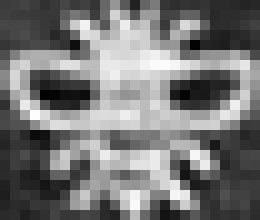}
    \end{minipage}\\
    \begin{minipage}{0.33\linewidth}
    \scriptsize{(b-i)}\\
    \includegraphics[width=0.9\linewidth]{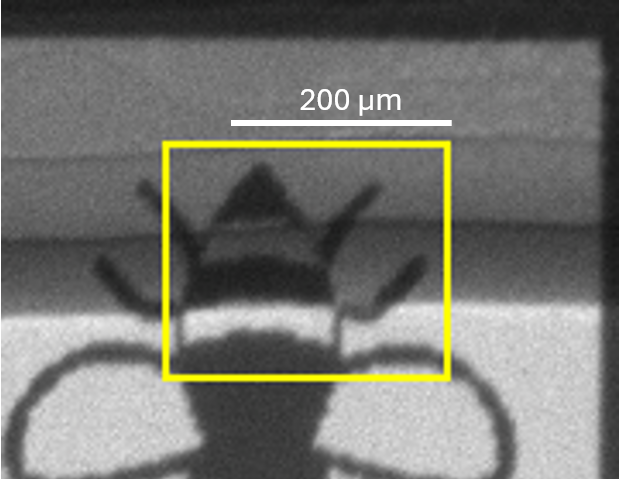}
    \end{minipage}%
    \begin{minipage}{0.33\linewidth}
    \centering
    \scriptsize{(b-ii)}\\
    \includegraphics[width=0.9\linewidth]{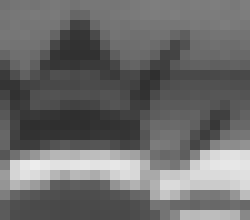}
    \end{minipage}
    \begin{minipage}{0.33\linewidth}
    \scriptsize{(b-iii)}\\
    \includegraphics[width=0.9\linewidth]{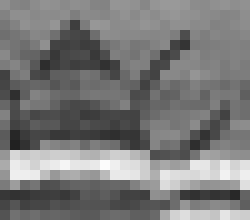}
    \end{minipage}%
    \caption{(a) image of a binary bee and (b) a multiple gray-level bee partially covered with several folded layers of Al tape. The recovered images (a-iii) and (b-iii) compare well relative to the expected images (a-ii) and (b-ii) respectively. The  $15 \times 15$ Huffman-like arrays was created from quaternary masks using pixels with pitch: (a) 10$\mu$m, (b) 20$\mu$m.}
    \label{fig:quaternary_results}
\end{figure}

Figure~\ref{fig:quaternary_results}(a-i) shows, in the highlighted region, a projected x-ray image of a binary object (a ``bee''). Figure~\ref{fig:quaternary_results}(a-ii) shows this region rescaled and binned to the same field-of-view and pixel size as the reconstructed image in Fig.~\ref{fig:quaternary_results}(a-iii). The image in Fig. ~\ref{fig:quaternary_results}(a-iii) shows the $P-N$ Huffman-like x-ray probe image obtained from the $15 \times 15$ mask (with 10 $\mu$m pixels) scanned over this object, after being deconvolved by the ideal Huffman-like array.

Figure~\ref{fig:quaternary_results}(b-i) shows an x-ray radiograph of another printed (``bee'') object, this time combined with several layers of aluminium folded in strips across the bee body to provide a range of object thicknesses. Figure~\ref{fig:quaternary_results}(b-ii) shows the highlighted region rescaled and binned to the same field-of-view and pixel size as the reconstructed image in Fig.~\ref{fig:quaternary_results}(b-iii). The image in Fig.~\ref{fig:quaternary_results}(b-iii) was obtained from the $P-N$ Huffman-like x-ray probe image obtained by scanning the $15 \times 15$ mask (with 20 $\mu$m pixels) over this ``gray'' object. The radiographs in Figs.~\ref{fig:quaternary_results}(a-ii) and \ref{fig:quaternary_results}(b-ii) compare well with the Huffman-like mask reconstructed images in Figs.~\ref{fig:quaternary_results}(a-iii) and \ref{fig:quaternary_results}(b-iii).  

\section{Discussion and future work} \label{sec:discussion}




The patterns used here for diffuse x-ray scanning probes based on Huffman-like arrays have been shown to have the ability to produce high-resolution images. In this case the resolution is dictated by the scanning step size and mask pixel size rather than the overall probe beam dimensions.
The tailored size and shape of these beams serve as a means to distribute the equivalent energy of a sharp probe over a much wider footprint. A broad, diffuse incident beam lessens the  potential radiation damage to the specimen and reduces thermal/structural changes.

We note that larger masks can spread the same energy of an equivalent beam much more thinly. However, they are harder to design while also retaining a small range of gray levels and require more complicated structures.  Also, masks with a large footprint require objects to be overscanned by at least the mask width outside the object edges to accurately recover internal detail out to the object edges. We observed that fabrication of masks with pixels of larger size (here 20 $\mu$m pixels rather than 10 $\mu$m) was shown to provide more uniform levels of transmission with more regular and sharply defined pixel edges.

Huffman-like masks, as differential filters, are signed operators. Masks that project positive beam intensities need to be split into positive, $P$, and negative, $N$, masks. The $P$ and $N$ masks are scanned individually over test objects, with $P$ and $N$ bucket values being collected individually (or as a group comprised of several mask regions, e.g. $[P,N/N,P]$). This duplication of scanning in both row and column directions increases the total imaging time, including the need to over-scan the areas around the object being imaged. However, scanning with the $[P,N/N,P]$ group of low-pass mask filters provides redundancy and choice in how to compute the differential Huffman-like mask, $P-N$, which can be composed in nine distinct ways (i.e. top-row differences $P_t-N_t$, bottom row $P_b-N_b$ or combinations $(P_t+P_b)-N_t$, $P_t-(N_t+N_b)$, etc.). The difference in timing where each of the four masks synchronise to scan the same point of the object also permits the possibility of correcting for variations in the brightness and uniformity of the incident illumination. 

Huffman-like masks with multiple transmission levels are shown here to perform better as imaging probes when used for the aperiodic scanning of objects than do binary masks based on perfect periodic arrays. Fabrication of masks with a smaller number of transmission levels is technically more robust, as  significantly fewer sequential deposition and removal operations are required. However, fabricating Huffman-like arrays with a larger number of transmission levels would have permitted significantly better autocorrelation metrics. For example, allowing the integer range of our $11 \times 11 $ Huffman-like arrays to increase to $\pm4$ from $\pm3$, would have improved the autocorrelation peak-to-sidelobe ratio to 26 from 20, the merit factor to 37 from 24, have flatter Fourier spectra (0.47 from 0.52), and lower condition number (1.24 from 1.33).

Binary masks are simpler to fabricate (with single Ta layer deposition) but require spatially larger pixels since they must be divided into subpixels over which the area of ``open'' transmission can be varied to accommodate a range of gray-level transmission of x-rays. The pixel area scales up with the range of transmitted intensities, further decreasing the spatial resolution of the probe. Orthogonal patterns of area filling of these sub-pixels are needed to reduce cross-pixel correlations. Any low-pass correlations between sub-pixels acts to reduce the delta-like property of the masks.

An alternative method to make gray-weighted $P$ and $N$ masks that uses a binary layer for transmission would be to switch pixel-sized micro-mirrors in or out for fractions of the total exposure time at each scan point. Such a method seems impractical for current x-ray imaging. However the Huffman-like masks developed here for x-rays are well-suited for optical imaging applications. A recently published special issue \cite{FlatopticsBeamShaping} addresses the shaping of optical beams at sub-wavelength scales to achieve a wide variety of probe objectives.

We note that a major constraint on the experimental work reported here was the time required between acquiring successive rasterscan data points. A combination of beam switching, translation stage movement and detector resetting meant about 3 seconds were required for each scan point. This limitation prohibited us scanning larger test objects. Even for a pinhole, we needed to over-scan the test objects by at least the width of the Huffman-like mask. Scanning a pinhole with a $32 \times 32$ mask in $[P,N/N,P]$ mode then took over an hour to complete; the small bee images took longer. Future experiments should reduce the scan-time by several orders-of-magnitude, using steps that match the actual beam exposure time used per point. 

In future experiments it may be useful to track the $(x, y)$ co-ordinates of the center of the probe pattern as the beam propagates along the $z$ direction. Similar work using Airy beams was reported in Ref. \cite{AiryLocalisation2020}. The masks designed in Ref. \cite{svalbe2021diffuse} may be useful here, as the delta-like mask planes also project as sharp delta-functions for several directions.

The design of $3D$ Huffman-like masks with a small range of intensity levels ($\pm3$) has been investigated briefly, with encouraging results for the construction of $11 \times 11 \times 11$ voxel arrays that have good autocorrelation metrics and condition number (an example is given in the  Supplementary Material). 

This diffuse-probe concept can be applied to existing scanning probe techniques such as x-ray fluorescence (XRF) imaging \cite{paunesku2006x}. Some consideration of how to employ the $P$ and $N$ masks separately is required. XRF microscopy involves a microfocused (pencil beam) x-ray probe scanned over a small (100 $\mu$m to 10 mm diameter) object and the x-rays fluorescing from the object are recorded with an energy-dispersive detector. The samples are typically biological and the focused beam can damage the sample. A Huffman-like diffuse probe would minimise radiation damage at the cost of a slightly larger scanning range. Since beam masking is used rather than beam focusing, the concept could also enable such a technique to be applied more easily with a laboratory x-ray source. How this concept could translate to more complicated scanning techniques such as ptychography \cite{pfeiffer2018x} is scope for future research. There are also future avenues to apply larger (high throughput) $[P,N/N,P]$ Huffman array masks in a static configuration as effective lenses in a coded-aperture context \cite{FenimoreCodedAperture1978, Coded-aperture-imaging-review}, to enable low-dose pinhole images via large flux, with sharp reconstruction. To this end, the pinhole results of Sec.~\ref{sec:Binary Huffman-like Mask Pinhole Images} and Sec.~\ref{sec:Quaternary Huffman-like Mask Pinhole Images} demonstrate the requisite stability to mask imperfections and slight misalignment issues.   

\section{Conclusion} \label{sec:conclusion}
We have successfully designed and fabricated a variety of Huffman-like masks, made as layers of tantalum deposited on a silica wafer. The masks, in sizes from $11 \times 11$ to $86 \times 86$ pixels, were fabricated with pixel widths that ranged from 8 $\mu$m to 20 $\mu$m. The masks were built to transmit discrete Huffman-like patterns of x-ray illumination with four distinct stepped levels of intensity.

The masks were exposed to near-monochromatic and uniform intensity incident x-ray beams. The radiographs of these masks, taken on a finely pixelated detector with 6.5 $\mu$m pixel pitch, bore close resemblance to the designed patterns. The autocorrelation of each reassembled mask x-ray image demonstrated a sharp delta-like pattern.

We acquired bucket signal images of various simple test objects using these masks. The bucket images from scanning the masks over a 20 $\mu$m pinhole aperture faithfully reproduced the designed array pattern. The deconvolved bucket x-ray images of the pinhole apertures also produced a delta-like correlation that confirmed that the pairs of low-pass bucket signals retained the broadband response expected from Huffman-like masks. The pinhole test images were a critical test of the mask practical functionality, as, with the pinhole illuminating only a tiny fraction of the area of the bucket detector, the image signal-to-noise is at a minimum.

The bucket images obtained using these fabricated masks on simple binary and stepped gray-level objects yielded realistic reconstructed high-resolution images of these objects after deconvolving the bucket images using the ideal integer arrays. This again confirms, conclusively, that the masks, as fabricated and imaged by the x-ray bucket signals, were able to replicate and conserve the delta-like properties of the Huffman-like arrays as designed.

Future work may consider applying similar masks to structure broad x-ray beams for x-ray fluorescence imaging applications and other existing scanning probe techniques.

\begin{acknowledgments}
AMK and DMP thank the Australian Research Council (ARC) for funding through the Discovery Project: DP210101312. AMK thanks the ARC and Industry partners funding the Industrial Transformation and Training  Centre for Multiscale 3D Imaging, Modelling, and Manufacturing: IC180100008. LR  thanks the ARC for funding through the Discovery Early Career Researcher Award DE240100006. IDS thanks Andrew Tirkel, Scientific Technology, Brighton, Australia, and Nicolas Normand with Jeanpierre Guedon, Ecole Polytechnique, Nantes, France, for collaboration on designing arrays with strong auto and weak cross correlation. This research was undertaken on the Micro-Computed Tomography beamline at the Australian Synchrotron, part of ANSTO, with beamtime and funding support allocated under grant application 20873. Andrew Stevenson, Benedicta Arhatari and Gary Ruben from MCT made themselves freely available and helpful throughout acquisition of the experimental x-ray imaging data. The authors acknowledge the facilities as well as the scientific and technical assistance of the Research and Prototype Foundry Core Research Facility at the University of Sydney, part of the NSW node of the NCRIS-enabled Australian National Fabrication Facility.
\end{acknowledgments}

\appendix
\section{Supplementary Material}

This Supplementary Material provides additional theoretical background and further technical detail on Huffman sequences and methods to compress their value range to form Huffman-like arrays. It also contains a more extensive description of the practical steps taken to fabricate Huffman-like masks by a precise pixel-wise deposition of patches of tantalum on a silica wafer. These masks were used to modify the transmitted intensity of x-ray beams to have Huffman-like profiles that were used as broad $2D$ scanning probes to image test objects. The diffuse pattern encoded by a Huffman-like intensity profile can be decoded by deconvolution to reconstruct a sharp image. That decoding property is made possible because of the delta-like autocorrelation of Huffman-like arrays. The advantage of scanning objects with a broad Huffman-like x-ray probe is to strongly reduce the rate of local energy deposition and hence minimise radiation damage.

\subsection{General Canonical Huffman Sequences Defined by Complex Polynomials}\label{sec:canonocalHuffmanSupp}

Huffman originally derived a sufficient and necessary criterion for constructing complex-valued canonical sequences, for which some integer forms, such as those based upon Lucas-Fibonacci polynomials \cite{HuntAckroyd1980, SvalbeTCI2020}, are known.  This subsection revises Huffman's construction, without any generalisation. 

Huffman defined a canonical sequence as the complex-valued coefficients $c_l$ of a polynomial $P(z) = \sum_{l=0}^{L-1}{c_l z^l}$, where $z_l$ denotes the $l^{\textrm{th}}$ integer power of a variable $z$ in the complex plane. For a canonical sequence of length $L-1$, Huffman defined a conjugate-reversed or complementary polynomial $Q(z) = \sum_{l=0}^{L-1}{\bar{c}_l z^{L-1-l}}$, where $\bar{c}_l$ is the complex conjugate of $c_l$. By construction, the roots of $P(z)$ have the same complex arguments (phase angles) as those of the conjugated $\bar{Q}(z)$ but the magnitudes of their roots are mutually reciprocal \cite{Huffman1962}. When expressed as a power series, the complex polynomial $P(z)\bar{Q}(z)$ has coefficients pertaining to the autocorrelation values of the sequence $c_l$. 

The canonical condition is that all autocorrelation elements need be zero, except the unavoidable ends with magnitudes $|\bar{c}_0 c_{L-1}|$ and peak value $A_0 = \sum_{l=0}^{L-1}{|c_l|^2}$ (sum of squared magnitudes). The product $P(z)\bar{Q}(z)$ then collapses to a quadratic in $z^{L-1}$, which has two roots.  Hence all roots of $P(z)\bar{Q}(z)$ must lie on either of two circles centered in the complex plane, with angles pertaining to the $(L-1)^\textrm{th}$ roots of unity, as is true for the roots of $P(z)$ and $\bar{Q}(z)$. Given the aforementioned reciprocal magnitudes between the roots of $P(z)$ and $\bar{Q}(z)$, this in turn means that these roots must occur on a circle of radius $R$ or $1/R$. There are $2^{L-1}$ choices one can make to place these roots around these circles, at equi-phase angles $\textrm{Arg}(z_l) = 2\pi l/(L-1)$, which produce (generally complex) Huffman sequences each having the same aperiodic canonical cross correlation, with peak value $A_0=|\bar{c}_0 c_{L-1}|(R^{L-1} - R^{-(L-1)})$. Hence we may write
\begin{equation}\label{eq:canonicalHuffmanCompute}
\mathcal{F}[H_{L}^s]_q =  c_{L-1}\prod_{l=1}^{L-1} (e^{2\pi i q/L}-R^{s_l}e^{2\pi i (l-1)/L}).
\end{equation}

The following comments apply in reference to Eq.~(\ref{eq:canonicalHuffmanCompute}). To numerically construct such a canonical Huffman sequence $H$ of length $L$ with elements $c_l$, one must choose a fixed radius $R$ (centred on $(0,0)$) for all complex roots $z_1$ to $z_{L-1}$ and then pick a set of signs $s_l \in \{-1,+1\}$, such that $z_l = R^{s_l}\exp(2\pi i l/(L-1))$. Real-valued $c_l$ are readily fixed by defining $z_l$ in the upper-half of the complex plane to have matching conjugated polynomial zeros in the lower half of the complex plane. The inverse discrete Fourier transform in Eq.~(\ref{eq:canonicalHuffmanCompute}) provides a convenient numerical algorithm for performing this computation. Due to potential underflow or numerical overflow for certain values of $R$, it is also advantageous to evaluate an exponential for a sum over logarithms, rather than to compute the series product directly. Since Eq.~(\ref{eq:canonicalHuffmanCompute}) represents a conversion between a set of polynomial roots and coefficients, the same generic computation can be used for non-canonical Huffman-like sequences, where the roots violate Huffman's criteria (for example when $|s_l| \ne 1$).

\subsection{Correlation and Fourier Properties of Huffman Sequences}\label{sec:SuppCorrelationFourierHuffman}
This section provides some examples of, and general properties for, particular integer Huffman sequences.

\begin{figure}
    \centering
    \includegraphics[width=0.8\linewidth]{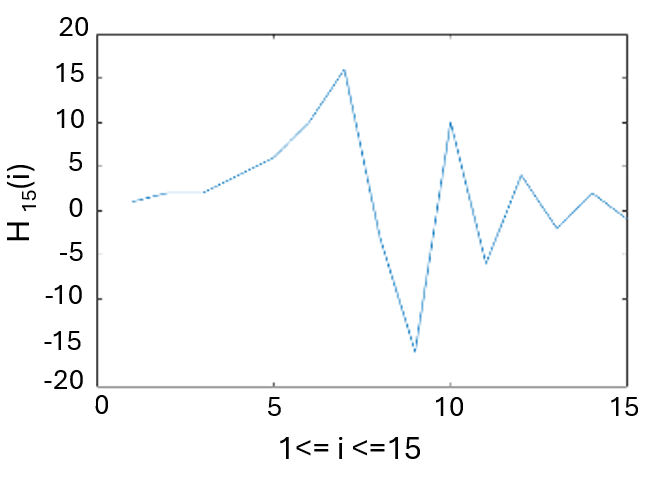}
    \caption{Canonical Huffman length $L=15$ integer sequence, $H_{15}$, the sequence values vary between $\pm16$. }
    \label{fig:canonical_Huffman}
\end{figure}

Integer values for the $1D$ Huffman sequence $H_{15}$ based on Lucas/Fibonacci series are:
$$H_{15} = [1,2,2,4,6,10,16,-3,-16,10,-6, 4,-2,2,-1]$$
as plotted in Fig.~\ref{fig:canonical_Huffman}. The autocorrelation, $H_{15}\otimes H_{15}$, of length $2\times15-1 = 29$, is 
$$H_{15}\otimes H_{15} = [-1, 0, 0, \cdots, 0, 843, 0, \cdots, 0, 0, -1],$$ 
which is as delta-like as possible under aperiodic conditions. 

Note the reflected left/right symmetry of the absolute values of the elements about the central element (here $-3$), with alternating sign changes for all the right elements. Any sequence with this reflected symmetry pattern of element values and signs ensures that every second autocorrelation value will be zero. To obtain zero off-peak autocorrelation values at \emph{all} but the (unavoidable) end elements requires a special choice for the sequence element values. Starting from the second element on the left, note that the next six values are twice the Fibonacci sequence, $2\times [1,1,2,3,5,8]$. Integer sequences built on this Lucas/Fibonacci pattern will have the maximum value as given by

\begin{equation}\label{eq:Huffman max range}
max(|H_L|) = \lfloor (2/\sqrt{5})\times\phi^{(L-3)/2} \rceil
\end{equation}
where $\phi$ is the golden ratio: $(1+\sqrt{5})/2$, and $\lfloor r \rceil$ denotes the integer round operation on real value $r$.

For $H_{31}$ there are 14 Fibonacci terms, giving the Huffman sequence a maximum value of $2\times377 = 754$. This result emphasizes why, for practical mask fabrication, strong but effective compression of the Huffman integer sequences is essential. For $2D$ arrays built using outer products, the maximum value for a non-compressed $31\times31$ integer Huffman array would be $754^2$.

The center term of the Lucas/Fibonacci integer sequences is also a Fibonacci term. In general it is always minus half the value of third Huffman term before the center. For our example $H_{15}$, the center term is $-(6/2) = -3$. 

In general, the sum of any sequence $S_L$, $\sum_{i=1}^{L}S_L(i)$, when squared, always equals the sum of \emph{all} the autocorrelation values. For Huffman sequences, the autocorrelation is non-zero only for the center term $A_0$ and the two end values. The autocorrelation (central) peak value $A_0$ is always the sum of all squared sequence values, $A_0 = \sum_{i=1}^L S_L(i)^2$. For Lucas/Fibonacci sequences, each end of the autocorrelation contributes an off-peak value of $-1\times+1 = -1$. Then, consistent with Parseval's Theorem:
$$\sum_{i=1}^L H_L(i)^2 = \left[\sum_{i=1}^L H_L(i)\right]^2 + 2.$$
For our example $H_{15}$ the sequence sum is $29$ and $29^2 = 841$. The autocorrelation peak for $H_{15}$, has $A_0 = 843$. 

The sequence sums also provide assurance that, after some range compression is applied, the compressed Huffman sequence $H_L^c$ remains closely Huffman-like, as then $$\left[\sum_{i=1}^L H_L^c(i)\right]^2 \approx \sum_{i=1}^LH_L^c(i)^2.$$ 

Not all integer Huffman sequences follow the Lucas/Fibonacci form. For example, at length $L = 11$ there are twin sequences
\begin{align*}
H_{11} &= [1,1,2,4,6,-1,-6,4,-2,2,-1], \\
H_{11,\textrm{twin}} &= [1,1,3,4,2,6,-7,-1,2,1,-1]
\end{align*}
for which, despite slightly different dynamic ranges, all the autocorrelation metrics are exactly the same (as is their aperiodic condition number, $\kappa$ = 1.0082). Another example of a different (but less practically useful) form of integer Huffman sequences is 
$$H_5 = [27,72,-24,8,-3],$$
with autocorrelation $$[-81,0,0,0,6562,0,0,0,-81].$$
This sequence has sum $80$. Note that $80^2 + 2\times81 = 6562$, the autocorrelation peak $A_0$.

\begin{figure}
    \centering
    \begin{minipage}{\linewidth}
    \centering
    \scriptsize{(a)}\\
    \includegraphics[width=0.9\linewidth]{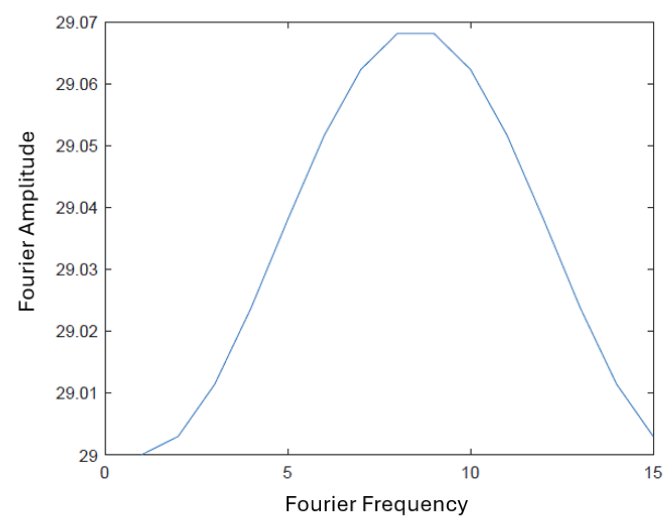}
    \end{minipage}\\
    \begin{minipage}{\linewidth}
    \centering
    \scriptsize{(b)}\\
    \includegraphics[width=0.9\linewidth]{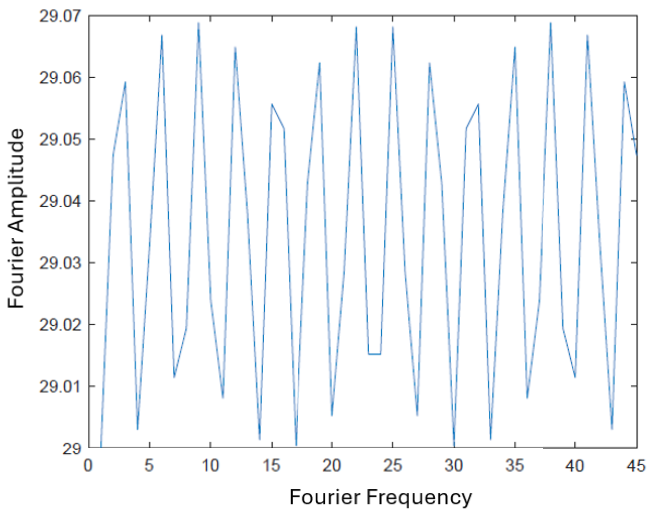}
    \end{minipage}%
    \caption{(a) Amplitude of FFT($H_{15}$) for the periodic array. (b) Amplitude of FFT($H_{15}$) for the aperiodic (zero-padded) array. The horizontal axis shows the Fourier frequency, $0-15$ for the periodic case, $0-45$ for the aperiodic case. The vertical axis shows the Fourier amplitude. The Fourier flatness $d_F = (29.07-29.00)/29.03 = 0.0024$ is the same for (a) and (b). }
    \label{fig:FFT_Huffman}
\end{figure}

The \emph{periodic} autocorrelation of any length $L$ Huffman sequence $H_L$, also has length $L$: 
$$(H_l \otimes H_L)_\textrm{periodic} = [0,\cdots,0,lr,A_0,rl,0,\cdots,0].$$ 
Now the product of the end terms ($l, r)$ of the sequence $H_L$ occurs in the autocorrelation at periodic shifts $\pm1$, with all the remaining cross-product terms summing to zero, as for the aperiodic case. The shape of the periodic and aperiodic (zero-padded for the aperiodic case) Fourier amplitudes, as shown for $H_{15}$ plotted in Fig.~\ref{fig:FFT_Huffman}, are characteristic for all real Huffman sequences.
The Fourier convolution theorem means the coefficients of $\mathcal{F}(H_L \otimes H_L)$ scale as $|\mathcal{F}(H_L)|^2$. The delta-like autocorrelation of $(H_L \otimes H_L)_{periodic}$ means there are only three terms that contribute to the periodic Fourier power spectrum: Fourier frequency $q = 0$ via the autocorrelation peak $A_0$, and from frequencies $q = \pm1$, arising from the product $lr$ of both end elements. For any real Huffman sequence, the inverted, single-period cosine shape for the \emph{periodic} Fourier spectral amplitude coefficients arises from the $q = \pm1$ terms that are added to the constant (mean value) contribution $q = 0$ from $\sqrt{A_0}$. The magnitude of the Fourier coefficients of $H_L$, for frequencies $q = 0, ..., L-1$, can then be written as: 
\begin{equation}\label{Fourierspectrumshape}
F(q) = \sum_{l=0}^{L-1} H_L(l) + \frac{|lr|}{\sqrt{A_0}} (1 - \cos(2 \pi q/L).
\end{equation}

As $\sum_{l=0}^{L-1} H_L(l)$ and $A_0 = \sum_{l=0}^{L-1} [H_L(l)]^2$ are both large relative to $lr$, the value for $d_F$, the Fourier flatness for Huffman sequences, is always close to zero (and hence any Huffman sequence autocorrelation is strongly delta-like, even for short sequences).

Note that, for any $2D$ Huffman array built from the outer product of a $1D$ sequence, the value of its $2D$ aperioidic autocorrelation peak, $A_0$, equals the \emph{square} of the peak autocorrelation value for the $1D$ sequence. Placed along each of the four edges of the autocorrelation are negated copies of $1D$ autocorrelation (that arise when the left/right and top/bottom edges of the array touch). All other off-peak autocorrelation values are zero. The autocorrelation for $3D$ Huffman arrays has the $2D$ correlation pattern on each of the six cube faces, with its $A_0$ values the cube of the $1D$ value (and all internal off-peak values being zero). 

Table \ref{tab:H7byH7aperiodic} shows that the aperiodic autocorrelation of the $2D$ array built from the $1D$ integer Huffman sequence $H_7$ is also optimally delta-like. In general, for Huffman sequence $H_L$, the metric value $R^0$ is the same in $1D$ as $2D$; the metric $M^f$ for $2D$ is always about half the $1D$ value. For $H_7, R^0 = 18$ for both $1D$ and $2D$ versions, whilst $M^f = 324/2$ in $1D$ and $\approx324/4$ in $2D$. 
\\
\begin{table}
    \centering
        \begin{tabular}{ccccccccccccc}
        1 & 0 & 0 & 0 & 0 & 0 & -18 & 0 & 0 & 0 & 0 & 0 & 1\\
        0 & 0 & 0 & 0 & 0 & 0 & 0 & 0 & 0 & 0 & 0 & 0 & 0\\
        0 & 0 & 0 & 0 & 0 & 0 & 0 & 0 & 0 & 0 & 0 & 0 & 0\\
        0 & 0 & 0 & 0 & 0 & 0 & 0 & 0 & 0 & 0 & 0 & 0 & 0\\
        0 & 0 & 0 & 0 & 0 & 0 & 0 & 0 & 0 & 0 & 0 & 0 & 0\\ 
        0 & 0 & 0 & 0 & 0 & 0 & 0 & 0 & 0 & 0 & 0 & 0 & 0\\
        -18 & 0 & 0 & 0 & 0 & 0 & 324 & 0 & 0 & 0 & & 0 & -18\\
        0 & 0 & 0 & 0 & 0 & 0 & 0 & 0 & 0 & 0 & 0 & 0 & 0\\
        0 & 0 & 0 & 0 & 0 & 0 & 0 & 0 & 0 & 0 & 0 & 0 & 0\\
        0 & 0 & 0 & 0 & 0 & 0 & 0 & 0 & 0 & 0 & 0 & 0 & 0\\
        0 & 0 & 0 & 0 & 0 & 0 & 0 & 0 & 0 & 0 & 0 & 0 & 0\\
        0 & 0 & 0 & 0 & 0 & 0 & 0 & 0 & 0 & 0 & 0 & 0 & 0\\
         1 & 0 & 0 & 0 & 0 & 0 & -18 & 0 & 0 & 0 & 0 & 0 & 1\\

        \end{tabular}
        \caption{The 2D aperiodic autocorrelation built from integer Huffman sequence $H_7$ is optimally delta-like.}
        \label{tab:H7byH7aperiodic}
\end{table}

\subsection{Generalised Reconstruction of Binarized Huffman Arrays}\label{sec:subPixelHuffmansSupp}

The array transpose $\intercal$ required to implement the separable scheme for binary sub-element arrays implies a slight modification to the de-correlation of Huffman array implementations as given in the main paper: 
\begin{equation}\label{eq:decorrSplit}
S_T =S_p-S_n=O \circledast P-O \circledast N = O\circledast H,
\end{equation}
%
in that each constituent sub-pixel array must be separately transposed in the de-correlating Huffman array (so that any pair of $Se_c$ and $Se_d$ is mutually geometrically orthogonal). For Huffman arrays defined by outer-products of Huffman sequences, one need only define the de-correlating Huffman array $H$ by a transpose of the entire array $H^\intercal$, as this is then the same as transposing each sub-element array. Denoting the transpose of individual sub-element arrays using a subscript $\intercal$, we can summarise the de-correlation as a minor variation on Eq.~(\ref{eq:decorrSplit}). 

Suppose two signals for positive and negative masks $Sb_p$ and $Sb_n$ are acquired with binary arrays $B_p$ and $B_n$, such that sub-element arrays $Se$ comprising $B_p-B_n$ correspond to elements of the Huffman array $H$. Denoting these measurements of a desired object $O$ as $Sb_p = O \circledast B_p$, $Sb_n = O \circledast B_n$, the binarized equivalent of Eq.~(\ref{eq:decorrSplit}) is
\begin{equation}\label{eq:decorrTranspose}
O \approx (Sb_p - Sb_n) \circledast (B_p - B_n)_{\intercal},
\end{equation}
where $(B_p - B_n)_{\intercal}=(B_p - B_n)^{\intercal}$ if and only if the corresponding Huffman array $H$ has transpose symmetry (as inherited from a lower-dimensional outer-product construction). 

\begin{figure}[htb!]
    \centering
    \includegraphics[width=0.8\linewidth]{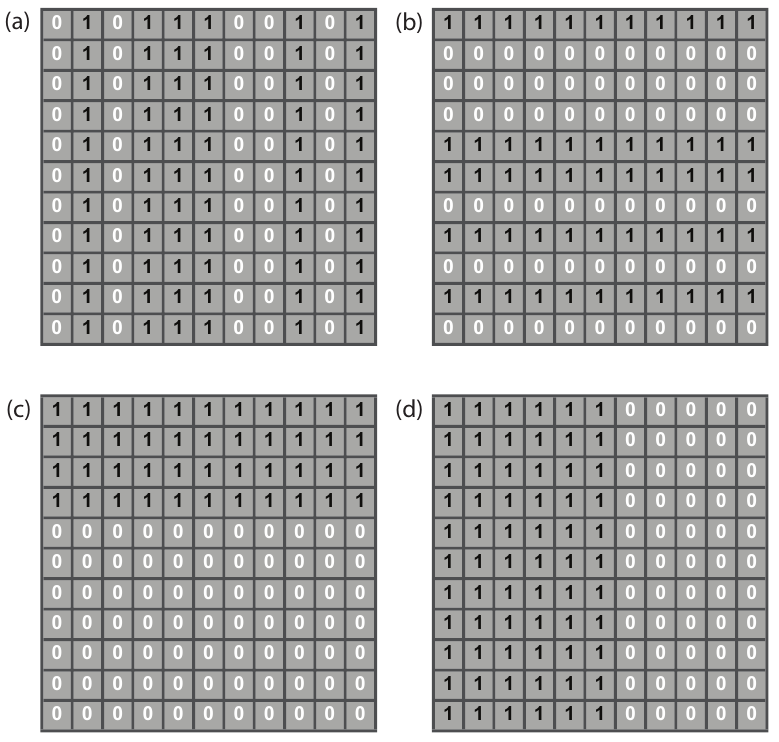}
    \caption{Sub-elements of a $2D$ Huffman array - the black/white ones/zeros represent 100\%/0\% transmission sub-elements respectively. (a) Randomised sub-element array for an original element value of six.  (b) Transpose of randomised sub-element array for decorrelation, for an original element value of five. (c) Contiguous randomised sub-element array for an original element value of four. (d) Contiguous randomised sub-element array for decorrelation, for an original element value of six.}
    \label{fig:subpixelHuffman}
\end{figure}

An example sub-element array for binary Huffman design is shown in Fig.~\ref{fig:subpixelHuffman}(a), corresponding to an original Huffman element with value six. A similar such transposed sub-element array required for de-correlation pertaining to a Huffman element value of five is shown in Fig.~\ref{fig:subpixelHuffman}(b).  
The simpler ``blocked'' design choices for practical fabrication are shown in Fig.~\ref{fig:subpixelHuffman}(c) and Fig.~\ref{fig:subpixelHuffman}(d). 

\begin{figure}[htb!]
    \centering
    \includegraphics[width=0.8\linewidth]{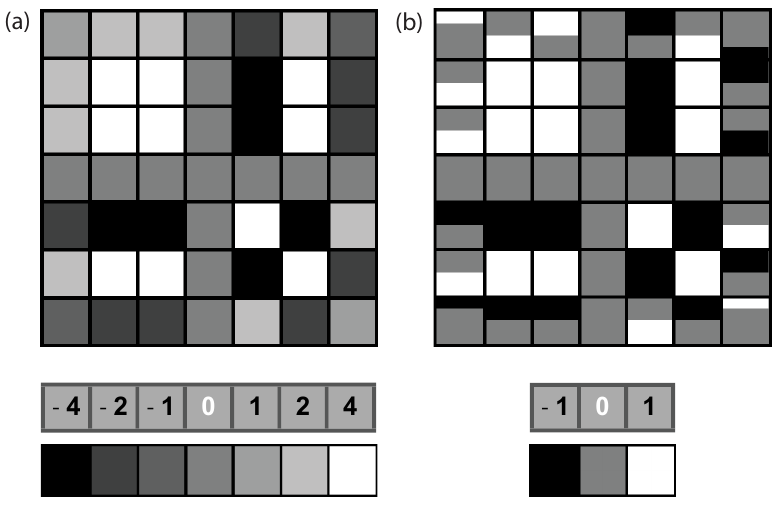}
    \caption{A canonical $7 \times 7$ Huffman array binarised. (a) An outer-product of a 7-element Fibonacci based sequence defines the $2D$ canonical Huffman array $H$, with gray levels in the range depicted by the legend at the bottom.  (b) Two binarised arrays subtracted from each other, namely $B_p-B_n$, as a ternary array with identical aperiodic auto-correlation to that of $H$ but which has $4 \times 4 = 16$ times as many elements.}
    \label{fig:HuffmanBinarised}
\end{figure}

As a complete example of the binarised Huffman transformation, Fig.~\ref{fig:HuffmanBinarised} depicts a canonical $7 \times 7$ sized $2D$ Fibonacci Huffman array stemming from an outer product of two identical $1D$ Hunt and Ackroyd \cite{HuntAckroyd1980} sequences with 7 elements each. The two signed binarised arrays $B_p$ and $B_n$, as given in the main paper, have been suitably superposed to represent the Huffman array in the ternary form $B_p-B_n$.  

\subsection{Optimising \texorpdfstring{$2D$}{2D} Huffman-like Arrays using Hybrid Reverse Monte Carlo}

Here we  provide an example of a HRMC simulation to optimise a $11 \times 11$ Huffman-like array, as per the algorithm outlined in the main paper. 
With the merit factor $M^f$ as an effective energy term, an example chi-squared $\chi^2$ that incorporates numerous other metrics 
such as spectral flatness $d^F$, fraction of zeros $f^z$, and off-peak correlation ratio $R^o$ can be written as:
\begin{align}
\begin{split}
\chi^2 = &(M^f - M^f_t)^2/M^f_w + (d^F - d^F_t)^2 /d^F_w + \\
&(f^z - f^z_t)^2 /f^z_w + (R^o - R^o_t)^2 /R^o_w,
\end{split}
\end{align}
where the subscript ``$t$” is the target value and ``$w$” is the weight.

As is typical for reverse Monte Carlo simulation, the individual weights are best empirically chosen. For the Huffman arrays produced in this work, this was done by gradually introducing constraints in several trial runs and checking that the running $\chi^2$ was suitably affected by new each new weight. Similar to topological constraints for atomic systems \cite{HRMCPorosityOpletal}, certain targets such as the fraction of zeros $f^z$ require comparatively strong weights (small weighting factors) in order to compete with all other target values. Interestingly trial simulations displayed ``cheating'' behaviour for $f^z$, whereby zeros would be concentrated at the borders, thereby emulating a smaller $2D$ Huffman array for which all other target metrics could be more readily achieved. To circumvent this issue, $f^z$ was measured with respect to the interior of the $2D$ array (with borders 20\% of the length).              

\begin{figure}[h!]
    \centering
    \includegraphics[width=\linewidth]{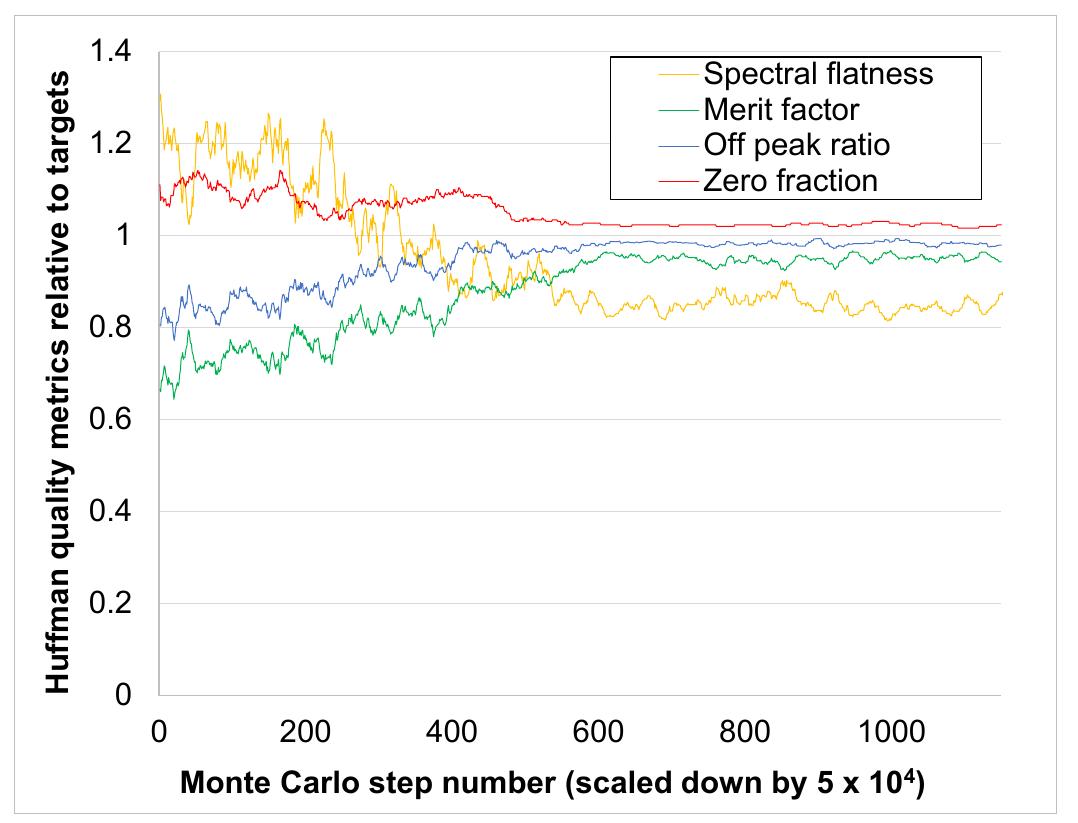}\\
\caption{Hybrid reverse Monte Carlo calculation to synthesize a $2D$ Huffman-like $11 \times 11$ array with integer-valued elements in $[-3,-2,-1,0,1,2,3]$.  All quality metrics shown have been normalised by their target values, such that a successful simulation would show all graphs meeting with unity values at the final Monte Carlo step.}
 \label{fig:HRMCHuffman}
\end{figure}

Figure~\ref{fig:HRMCHuffman} shows an example HRMC simulation for an $11 \times 11$ Huffman array, for integer elements ranging from -3 to +3. During the simulated annealing stage, the dimensionless temperature was linearly ramped from an initial 300 value down to 20 at the mid-point of the x-axis (step number $3 \times 10^7$), where-after it was held at 20 until the end of the simulation. For the simplicity of presentation for this particular example, the Monte Carlo step size was not weighted by $kT$, nor was the range of integer values. Similarly, unlike the HRMC simulations for the Huffman-like arrays used in experiment, the maximum gray level 
was maintained at value 3 for the entire simulation (i.e. the degree of quantising was not gradually changed as another form of simulated annealing). 

Auto-correlation quality metrics including the spectral flatness, merit factor and off peak ratio were optimised to their dimensionless target values of 0.8, 15, and 24 respectively, while other metrics were not optimised. At Monte Carlo step number $6 \times 10^7$ (1200 on the scaled x-axis), these appear respectively as the lowest, second lowest and second highest fluctuating curves. The highest such curve in Fig.~\ref{fig:HRMCHuffman} is the fraction of zero values in the $2D$ array, which was optimised to a target value of 0.25.  

As evident in Fig.~\ref{fig:HRMCHuffman}, all quantities eventually met their target values, tending toward unity, with the exception of the spectral flatness $d^F$. This metric was flatter than was thought possible, leading to a more canonical Huffman array. For small $2D$ arrays such as this example, it is possible that only certain sets of metrics are commensurate and it is an open question as to how much these can be simultaneously optimised. The final (randomly fluctuating) relaxed coefficients of the complex polynomial for this simulation is the Huffman-like sequence 
\begin{align}
\begin{split}
t_{11} = [&0.5143, 0.6039, 0.2587, 0.3291, 1.3959, 1.3873, \\
&-0.7717, -1.2099, 0.8503, -1.1455, 1.0000].
\end{split}
\end{align}
The $2D$ HRMC Huffman example from which the performance metrics were computed was given by the integer rounded outer product of $t_{11}$.

\subsection{Image Reconstruction: Deconvolution and Deblurring}\label{sec:deconvolve and deblur}
The following section and sub-sections describe further optimisation processing that helped to fine-tune the performance of Huffman-like arrays as the range of element values was compressed relative to the original Huffman array. Arrays that have fewer and smaller off-peak autocorrelation values have better autocorrelation metrics but also yield smaller image reconstruction errors (such as the mean squared error per pixel) when applied to arbitrary test data.

The 2D and 3D Huffman-like masks, $H$, were tested in computer-simulated conditions by convolving digital test data, $I$, with the designed Huffman-like array imprints to produce ``bucket images'', $B(x, y, z) = I \otimes H$, where each bucket value sums all of the intensities that fell under the beam footprint for each translation $(x, y, z)$ across the test data (including an over-scan region that extends beyond the object borders to allow for coverage by the finite size of the beam probe).

The images are reconstructed by cross-correlating their bucket image $B$ with the Huffman-like mask $H_{\intercal}$. The off-peak autocorrelation values are known to be small (relative to the correlation peak) and sparse (i.e. there are many autocorrelation zeros). A simple deblurring can be achieved by iterative subtraction of the known (and unwanted) contribution of each off-peak autocorrelation contribution to the original deconvolved image. A scaled and shifted copy of the reconstructed image is subtracted for each non-zero off-peak autocorrelation location. This deblurring procedure also works to reduce the image reconstruction errors that arise from the unavoidable end or edge contributions of the canonical Huffman sequences and arrays.

\subsubsection{Deblured Image Reconstruction after $2D$ Huffman-like Deconvolution}\label{sec:deconvolve results}
For the best-performing $11\times 11$ Huffman-like masks with element range $\pm3$, convolved test data were able to be deconvolved and deblurred to reconstruct images of the test data with mean absolute errors of less than 1 gray level per pixel for 8-bit test data. The data ranged from $191 \times 191$ flat disc images to high contrast pictorial image data with near-flat discrete Fourier coefficients.

The better masks showed deblurring that converged to a stable mean error (of 0.1 gray levels per pixel that remained near-constant after 10 and further deblur cycles). Some larger masks had reconstruction errors that converged to a minimum after a few deblur cycles but that slowly worsened with further deblurring.

Interestingly, whilst the image reconstruction errors were smaller for arrays with better correlation metrics, that was not strictly always the case for all test data. The transpose symmetric arrays also perform slightly differently on the same test data when rotated by $90^\circ$.

\begin{figure}
    \centering
    \includegraphics[width=\linewidth]{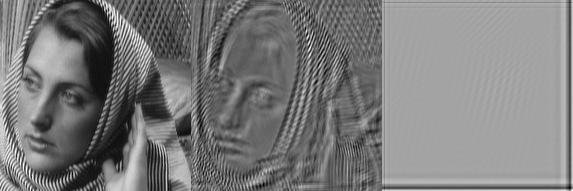}
    \caption{Left: Original $191 \times 191$ portion of 8-bit ``Barbara'' data (zero-padded before array convolution). Centre: deconvolved image reconstruction errors, 0 deblur cycles. Right: image reconstruction errors, 10 deblur cycles. All images shown as max = white = 255, min = black = 0. For actual error image values, see the entries for 0 and 10 deblur cycles in Table~\ref{fig:recon_2Deg_data}.}
    \label{fig:recon_2Deg_barbara}
\end{figure}

The results of image reconstruction from data generated by convolution with an $11 \times 11$ Huffman-like array are presented in Fig.~\ref{fig:recon_2Deg_barbara}. For this 2D example, a $191 \times 191$ pixel subset of the ``Barbara'' image is used as the input image. The reconstruction process involves deconvolution and iterative deblurring. Improvement in the quality of reconstructed images over the first 10 deblurring cycles is presented in Table~\ref{fig:recon_2Deg_data}.

\begin{table}
    \centering
    \begin{tabular}{ccccc}
        \toprule
       D  & $mabs$ & $max$ & $min$ & dc error\\
        \midrule
        \textbf{0} & \textbf{5.1176} & \textbf{35.0819} & \textbf{-28.5089} & \textbf{2.8642}\\
        1 & 1.5412 & 10.7602 & -11.3396 & -0.1160\\
        2 & 0.6741 & 5.9158 & -5.3461 & 0.0146\\
        3 & 0.3636 & 3.1774 & -2.8311 & 0.0172\\
        4 & 0.2280 & 1.7656 & -2.2120 & 0.0242\\
        5 & 0.1606 & 1.1072 & -1.8740 & 0.0246\\
        6 & 0.1239 & 1.3733 & -2.1798 & 0.0255\\
        7 & 0.1037 & 1.1051 & -2.0187 & 0.0256\\
        8 & 0.0930 & 1.2645 & -2.1303 & 0.0257\\
        9 & 0.0879 & 1.1606 & -2.0614 & 0.0257\\
        \textbf{10} & \textbf{0.0855} & \textbf{1.2236} & \textbf{-2.1055} & \textbf{0.0257}\\
        \bottomrule
    \end{tabular}
    \caption{$191 \times 191$ pixel image reconstruction errors after convolution of data by Huffman-like $11 \times 11$ mask and applying $D$ deblur cycles.The error measures are, left to right, mean absolute, maximum, minimum, and mean values.}
    \label{fig:recon_2Deg_data}
\end{table}

\subsubsection{Deblurred Image Reconstruction after $3D$ Huffman-like Deconvolution}\label{sec:3D deconvolve results}


\begin{figure*}
    \centering
    \includegraphics[width=\linewidth]{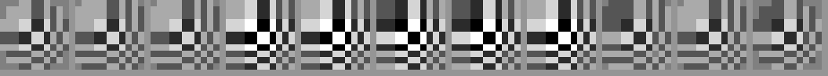}
    \caption{3D Huffman-like array shown as 11 slices, each $11 \times 11$ pixels, top to bottom in $z$. This cube of 1331 voxels has full 3D transpose symmetry. Black = -3, white = +3. Array sum $ = 35$, $RMS = 1.65$, $MAV = 1.48$. Autocorrelation metrics: $R^o = 20.60$, $M^f = 12.27$, $d^f = 0.94$, the autocorrelation peak value is 3625, off-peak values range from -176 to +117, 5214 of the 9261 autocorrelation values equal zero, the aperiodic condition number $\kappa = 1.49$}
    \label{fig:3D_Huffman_H11}
\end{figure*}


\begin{table}
    \centering
    \begin{tabular}{ccccccc}
        \toprule
       Slice  & $R^0$ & $M^f$ & $d^f$ & $\kappa$ & tp & sgn\\
        \midrule
        1 & 12.57 & 6.19 & 1.26 & 1.85 & 0 & 0\\
        2 & 10.67 & 9.90 & 0.64 & 1.60 & 0 & 0\\
        3 & 10.67 & 9.90 & 0.64 & 1.60 & 0 & 0\\
        4 & 19.27 & 21.41 & 0.56 & 1.31 & 0 & 0\\
        5 & 19.27 & 21.41 & 0.56 & 1.31 & 0 & 0\\
        6 & 19.27 & 21.41 & 0.56 & 1.31 & 0 & 0\\
        7 & 19.27 & 21.41 & 0.56 & 1.31 & 0 & 0\\
        8 & 19.27 & 21.41 & 0.56 & 1.31 & 0 & 0\\
        9 & 10.67 & 9.90 & 0.64 & 1.60 & 0 & 0\\
        10 & 10.67 & 9.90 & 0.64 & 1.60 & 0 & 0\\
        11 & 12.57 & 6.19 & 1.26 & 1.85 & 0 & 0\\
        \bottomrule
    \end{tabular}
    \caption{The aperiodic autocorrelation metrics for each of the 11 ($x$ or $y$ or $z$ axis) 2D planes of this 3D array. $\kappa$ = aperiodic slice condition number, tp = 0 if the plane is transpose symmetric, sgn = 0 if slice($i$) = slice($12-i$) for $i$ even, and if slice($i$) = -slice($12-i$) for $i$ odd.}
    \label{tab:3D aperiodic autocorrelation slice metrics}
\end{table}

\begin{figure}
    \centering
    \includegraphics[width=\linewidth]{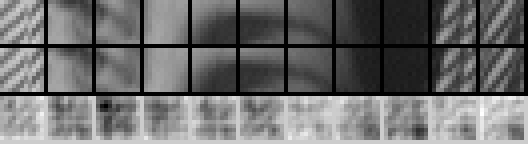}
    \caption{Reconstruction errors for 3D Huffman-like $11 \times 11 \times 11$ array. Top row: 11 slices of $11 \times 11$ original data, portions of Barbara image. Middle row: the reconstructed image slices from zero-padded data. Bottom row: Image reconstruction errors in each slice after 5 deblur cycles. All images are shown scaled max = white = 255, min = black = 0. Actual reconstruction errors values are printed in Table~\ref{table:recon_3Deg_data} The smallest reconstruction errors occur here after four deblur cycles.}
    \label{fig:recon_3Deg_barbara}
\end{figure}

The results of 3D image (or volume) reconstruction from data generated by convolution with an $11 \times 11 \times 11$ Huffman-like array are presented in Fig.~\ref{fig:recon_3Deg_barbara}. For this 3D example, 11 slices of $11 \times 11$ pixel subsets of the ``Barbara'' image are used to synthesize the input volume. The reconstruction process involves deconvolution and iterative deblurring. Improvement in the quality of reconstructed images over the first 5 deblurring cycles is presented in Table~\ref{table:recon_3Deg_data}.

\begin{table}
    \centering
    \begin{tabular}{ccccccc}
        \toprule
        cycle & $min$ & $mabs$ & $max$ & DC\\
        \midrule
        0 & -0.24 & 24.40 & 67.25 & 24.40 \\
        1 & -1.81 & 10.40 & 23.61 & 10.39 \\
        2 & -6.49 & 3.42 & 10.85 & 3.01 \\
        3 & -6.33 & 1.69 & 5.86 & 0.61 \\
        4 & -7.57 & 1.48 & 3.26 & -0.95 \\
        \textbf{5} & \textbf{-8.29} & \textbf{1.71} & \textbf{2.70} & \textbf{-1.49} \\
        \bottomrule
    \end{tabular}
    \caption{Zero-padded $11 \times 11 \times 11$ data image reconstruction errors after convolution of data by Huffman-like $11 \times 11 \times 11$ mask for the first 5 deblur cycles. The errors are, left to right, the minimum, mean absolute, maximum and mean. Note here the mean absolute error per pixel is least after four deblur cycles.}
    \label{table:recon_3Deg_data}
\end{table}

\subsection{Autocorrelation of Binary $P$ and $N$ X-ray Images as Recombined Huffman-like Masks}
Validation of the x-ray images taken of the quaternary Huffman-like masks (as $P$ and $N$ mask images combined to form the Huffman-like mask) was presented in the main paper. The multi-level transmission required for the quaternary masks was more difficult to fabricate. This section presents x-ray image validation results for the larger but much simpler to fabricate binary masks.

\begin{figure}
    \centering
    \includegraphics[width=0.8\linewidth]{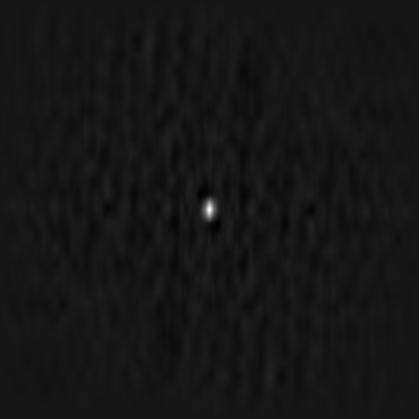}
    \caption{Image of the autocorrelation of the fabricated binary $15\times15$ mask. Note the peak shape is wider vertically than horizontally. Small, similarly elongated  off-peak entries are also visible scattered across the image.}
    \label{fig:binary_mask_expt_autocorr}
\end{figure}

 The images obtained from the $[P,N/N,P]$ mask regions were used to reassemble an image of the signed Huffman-like $15\times15$ mask. The autocorrelation of the reassembled mask image, shown in  Fig.\ref{fig:binary_mask_expt_autocorr}, shows that the reassembled mask has retained its delta-like property.

\begin{figure}
    \centering
    \begin{minipage}{0.5\linewidth}
    \centering
    \scriptsize{(a)}\\
    \includegraphics[width=\linewidth]{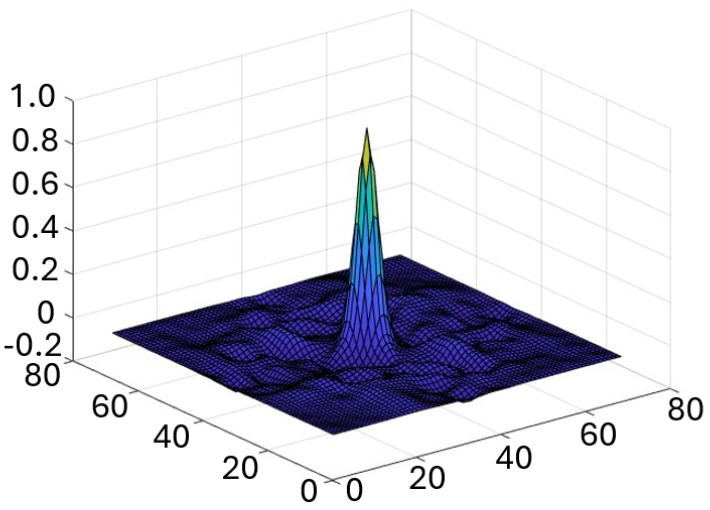}
    \end{minipage}%
    \begin{minipage}{0.5\linewidth}
    \centering
    \scriptsize{(b)}\\
    \includegraphics[width=\linewidth]{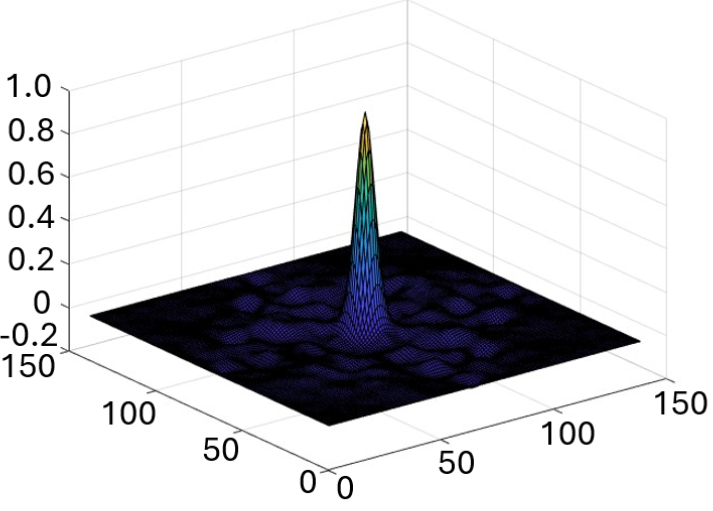}
    \end{minipage}\\
    \begin{minipage}{0.5\linewidth}
    \scriptsize{(c)}\\
    \includegraphics[width=\linewidth]{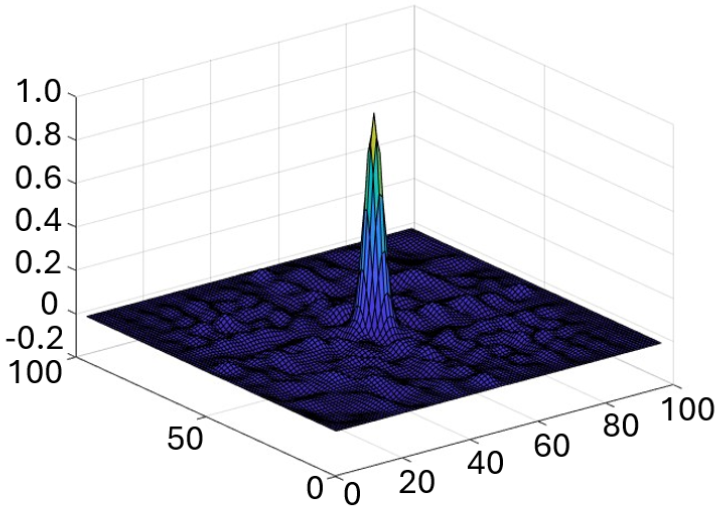}
    \end{minipage}%
    \begin{minipage}{0.5\linewidth}
    \centering
    \scriptsize{(d)}\\
    \includegraphics[width=\linewidth]{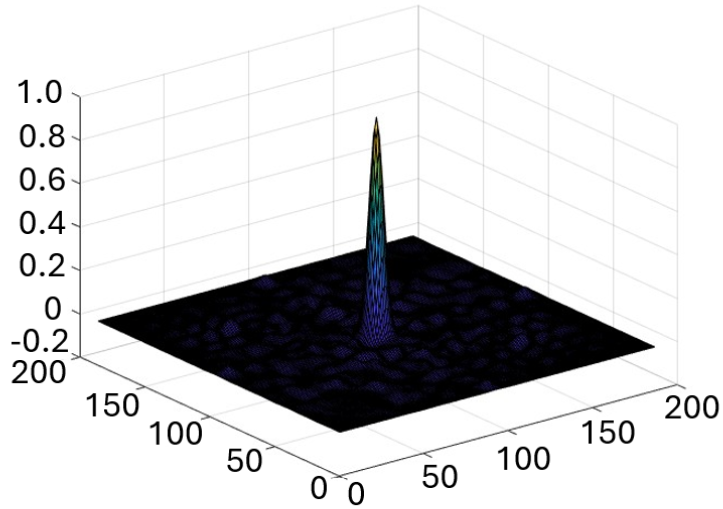}
    \end{minipage}%
    \caption{Normalised surface plots for the autocorrelations of detector-measured x-ray images of the binary fabricated masks: (a) $11\times11 \pm3$ gray-levels, $10$ micron pixels, (b) $20$ micron pixels. Real mask images  $15\times15 \pm3$ gray-levels with (c) $10$ micron pixels and (d) for $20$ micron pixels. The vertical scales have been normalised to value $1$.}
    \label{fig:quaternary_expt_autocorr}
\end{figure}

Figure \ref{fig:quaternary_expt_autocorr} shows the  autocorrelations of the four detector images obtained for the two $11\times11$ and the two $15\times15$ reassembled masks. The fabricated $P$ and $N$ masks are both intrinsically low-pass filters. Composed as $P-N$, the masks become broad-band filters with autocorelations that are  delta-like. The detector images formed by the projected x-ray mask $P$ and $N$ regions had sizes $35\times35, 70\times70, 49\times49, 94\times94$, hence the autocorrelation peak widths for those images are slightly broader than the single pixel width expected for $11\times11$ and $15\times15$ arrays. 

\subsection{Mask Fabrication}
The binary mask design consists of transparent and opaque parts. In terms of fabrication, the transparent part is the substrate, which can be glass or $\textrm{SiO}_2$, while the opaque part is made of a material that is sputtered and patterned onto the substrate. Ideally the sputtered material should be thick enough to block almost all the x-rays at a specific photon energy. However, there are practical limitations of the sputtering process itself. The maximum achievable film thickness using the sputtering technique is approximately 4 $\mu$m to 5 $\mu$m, which remains challenging. The reason is that such a thick film experiences a high level of stress that can result in cracking, delamination, or other mechanical issues. In addition, there are other issues such as non-uniformity, target erosion, and process instability as the thickness increases to more than a few microns. Thus, the maximum thickness of the mask material is limited to 5 $\mu$m.

\begin{figure}[h!]
    \centering
    \includegraphics[width=\linewidth]{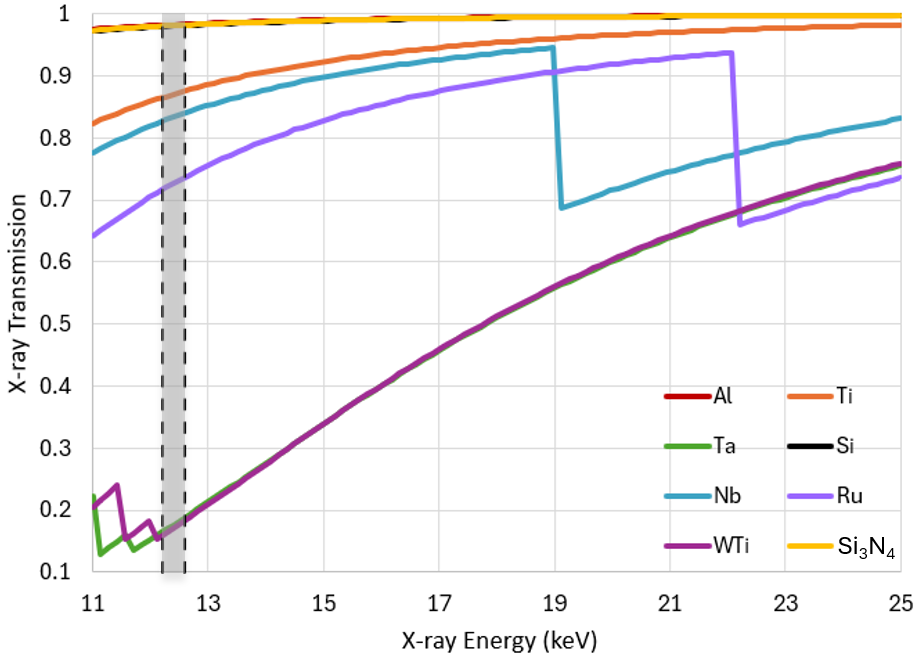}\\
\caption{X-ray transmissions through 5 $\mu$m thickness of selected materials at a range of photon energies from 11 keV to 25 keV. The gray area between the dashed lines shows the 3$\%$ energy bandpass at the selected energy of 12.4 keV.}
 \label{fig:Tr plot}
\end{figure}

To select a suitable mask material, a study was conducted to calculate x-ray transmission through 5 $\mu$m of various materials available for the sputtering tool at the Research and Prototype Foundry (RPF) at The University of Sydney, Australia, where the masks were fabricated. This analysis covered a range of photon energies from 11 keV to 25 keV as shown in Fig.~\ref{fig:Tr plot}. The available materials included Al, Ti, Ta, Si, Nb, Ru, WTi, and $\textrm{Si}_3$$\textrm{N}_4$. Based on the x-ray transmission plot in Fig.~\ref{fig:Tr plot}, WTi and Ta were selected as appropriate materials because of having relatively low x-ray transmissions over the selected photon energies. Among these two materials, Ta was chosen as the material for sputtering purposes due to two main reasons: firstly, Ta films tend to offer better uniformity and adhesion compared to WTi films, especially at thicker deposition levels. Secondly, optimising sputtering parameters and conditions are generally straightforward for a single material like Ta compared to complex alloy compositions like WTi. 

As shown in Fig.~\ref{fig:Tr plot}, a 5 $\mu$m thickness of Ta has a minimum transmission of approximately 17.5$\%$ at 12.4 keV photon energy. Note that 12.4 keV was selected rather than 12 keV to avoid the L edge of the Ta and to consider the energy bandpass of 3$\%$ at Micro-Computed Tomography (MCT) beamline \cite{arhatari2023micro} of the Australian Synchrotron where the experiments were performed. The final step before starting the fabrication process was to choose a substrate, which acts as the transparent parts of the masks. $\textrm{SiO}_2$ was selected as the substrate due to its widespread availability and common usage. Additionally, its transmission under 12.4 keV is higher than that of Si, another commonly used substrate. The higher transmission of the substrate can improve the contrast of the acquired experimental images. 

\subsubsection{Binary Masks} \label{sec:fabProcess_binary_masks}
\begin{figure}[h!]
    \centering
    \includegraphics[width=\linewidth]{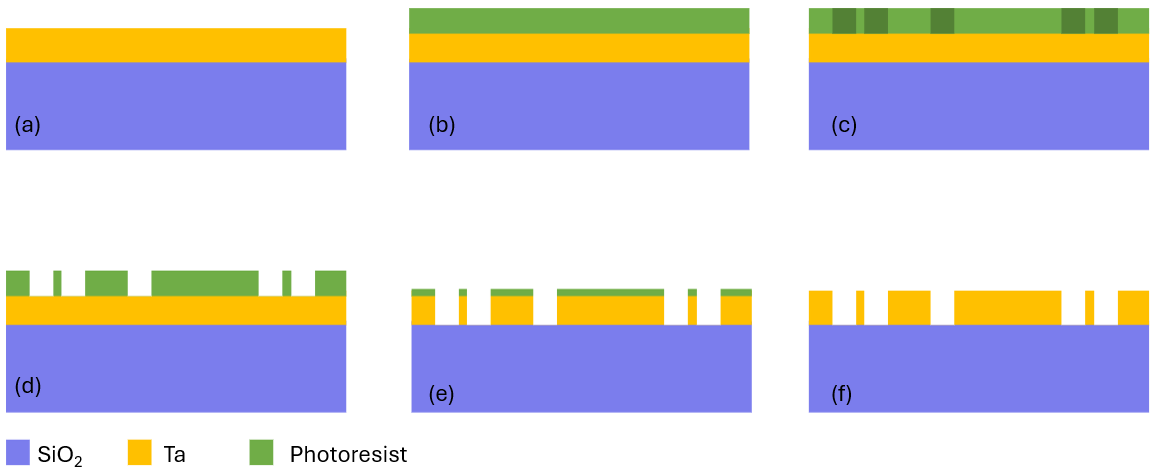}\\
\caption{Schematic of the fabrication process for the binary masks: (a) sputtering a tantalum layer on the $\textrm{SiO}_2$ substrate, (b) spin-coating a photoresist, (c) writing the mask patterns onto the photoresist layer, (d) developing the photoresist, (e) etching Tantalum, and (f) removing the remaining photoresist.}
 \label{fig:Fab Process Binary}
\end{figure}

Binary masks were fabricated in six steps as schematically depicted in Fig.~\ref{fig:Fab Process Binary}. First, a 6-inch $\textrm{SiO}_2$ wafer, the substrate, was cleaned with acetone and isopropanol (IPA). Then, the wafer was placed in the sputtering chamber for the sputtering process using an AJA ATC-2200 Sputtering Deposition System. The deposition rate was approximately 8.5 nm per minute. The sputtering time was about 10 hours to achieve nearly 5 $\mu$m thick Ta on the $\textrm{SiO}_2$ wafer. A low deposition rate was necessary to minimise the film stress. After the sputtering process, the 6-inch wafer was coated with a photoresist and then cut into 2 cm $\times$ 2 cm pieces using a Dicing Saw (ADT) machine. The photoresist was used to protect the Ta layer during the dicing process, and it was removed (with acetone) afterwards. The rest of the fabrication process was conducted on individual 2cm $\times$ 2cm pieces. 

After preparing the Ta pieces (see Fig.~\ref{fig:Fab Process Binary}(a)), a lithography process was applied in three steps: 1- spin-coating a photoresist onto the substrate (Fig.~\ref{fig:Fab Process Binary}(b)), 2- patterning (Fig.~\ref{fig:Fab Process Binary}(c)), and 3- developing the photoresist (Fig.~\ref{fig:Fab Process Binary}(d)). Ma-P 1275G (Micro resist technology) was selected as the photoresist because it provides a relatively thick layer (more than 3 $\mu$m), which is required for the subsequent etching process. The photoresist was spin coated onto the Ta layer at the maximum velocity of 3000 rpm for 50 seconds and then baked at 105 $^\circ$C for 120 seconds. At this stage the wafer was ready for patterning. A maskless aligner (Heidelberg MLA100) was used to write binary Huffman-like patterns into the wafer. The dose and the defocus parameters were set to 1000 $mJ/cm^2$ and 0 respectively. These parameters were chosen from a dose study that we conducted on a few pieces of the same wafer. 

\begin{figure}[h!]
    \centering
    \includegraphics[width=\linewidth]{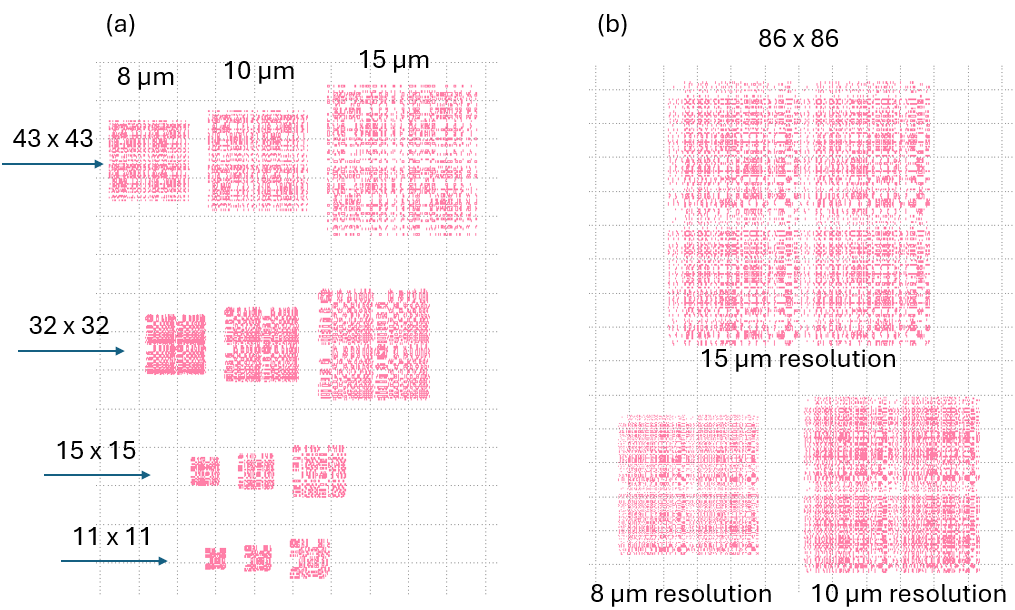}\\
\caption{Layout of some of the binary Huffman-like masks with different sizes and resolutions.}
 \label{fig:Ternary design}
\end{figure}

Layouts for the binary mask designs used for the maskless aligner are shown in Fig.~\ref{fig:Ternary design}. Figure.~\ref{fig:Ternary design}(a) shows 12 binary masks with four different sizes (i.e. $11\times11$, $15\times15$, $32\times32$, and $43\times43$) and three resolutions (i.e. 8$\mu$m, 10$\mu$m, and 15$\mu$m). These patterns were written on a 2cm $\times$ 2cm piece of the wafer. The $86\times86$ binary Huffman-like mask patterns, shown in Fig.~\ref{fig:Ternary design}(b), were written into another piece of wafer. The $86\times86$ binary mask were also fabricated with 8$\mu$m, 10$\mu$m, and 15$\mu$m resolutions. After writing the patterns, the wafers were developed using an AZ 726 MIF Developer for 7 minutes, followed by rinsing with deionised (DI) water and drying with nitrogen gas. The next step was to etch through the Ta layer using a Reactive Ion Etcher (RIE), as shown in Fig.~\ref{fig:Fab Process Binary}(e). The wafers were placed in the RIE chamber (Plasmatherm Vision) and etched for 50 minutes. This time was sufficient to etch the Ta layer completely and reach to the $\textrm{SiO}_2$ layer. The RIE recipe was a combination of $\textrm{SF}_6$, $\textrm{CF}_4$, $\textrm{CHF}_3$, and $\textrm{O}_2$ gases, which we optimised to deep etch the Ta layer. In the last fabrication step, the remaining photoresist was removed by acetone and then the wafer was rinsed and dried by DI water and nitrogen gas respectively (see Fig.~\ref{fig:Fab Process Binary}(f)). 

\subsubsection{Quaternary Masks} \label{sec:fabProcess_Quaternary}
Quaternary masks, as explained in the text, have four levels $(0, 1, 2, 3)$. Each level transmits an x-ray beam with steps of increasing intensity. Multiple degrees of x-ray transmission through a mask can be achieved by varying the thickness of the mask material. Pixels at each level of the quaternary mask can be fabricated with a specific thickness to provide the required level of x-ray transmission. As for the binary mask, Ta was chosen as the mask material. Given the maximum achievable thickness of 5 $\mu$m using the sputtering technique, as discussed in the previous section, we can estimate the minimum x-ray transmission through our mask. As shown in Fig.~\ref{fig:Tr plot}, the minimum transmission through a 5 $\mu$m Ta is approximately 17.5$\%$ at 12.4 keV. This is the transmission through level 0. Based on this minimum transmission, the transmissions through levels 1, 2, and 3 are calculated as 45$\%$, 72.5$\%$, and 100$\%$ respectively. Having the x-ray transmissions (T), the thickness of each level can be measured as
\begin{equation}
t = -ln(T)/\mu,
\end{equation}
where $\mu$ is the linear attenuation coefficient. Using this formula the thicknesses of the levels 1, 2, and 3 were calculated as approximately 2.2 $\mu$m, 0.88 $\mu$m, and 0 respectively.

\begin{figure}[h!]
    \centering
    \includegraphics[width=\linewidth]{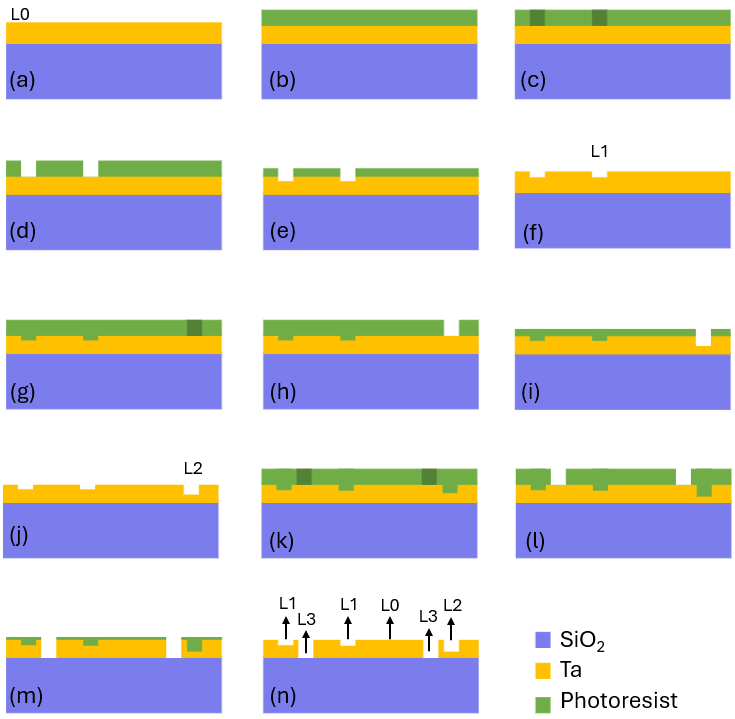}\\
\caption{Schematic of the fabrication process for the quaternary masks.(a) Ta film deposition on a $\textrm{SiO}_2$ substrate. (b)-(d) first lithography process, which includes (b) spin-coating a photoresist on the Ta layer, (c) patterning the photoresist, and (d) developing the photoresist. (e) first etching process to define level one of the quaternary masks. (f) removing the remaining photoresist. (g)-(h) second lithography, and (i) etching processes to define level two of the quaternary masks. (j) photoresist removal. (k)-(l) third lithography and (m) etching processes to define level three of the quaternary masks. (n) final result after removing the remaining photoresist.}
 \label{fig:4level Fab Process}
\end{figure}

\begin{figure}[h!]
    \centering
    \includegraphics[width=\linewidth]{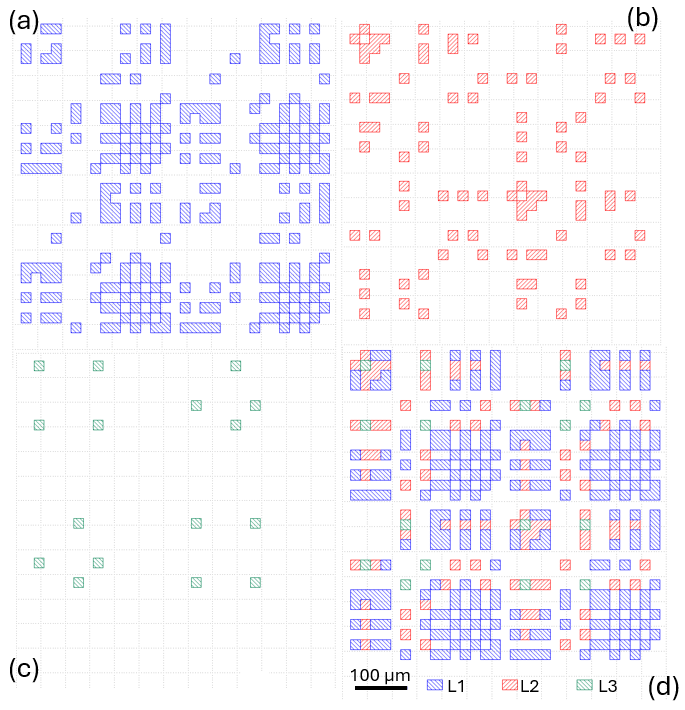}\\
\caption{An example of the CAD file for a quaternary mask. Layers 1 (a), layer 2 (b), and layer 3 (c) of the $15\times15$ quaternary mask design with 20 $\mu$m resolution. (d) is the combination of all layers. The scale bar is for all the images.}
 \label{fig:Huffman layout}
\end{figure}

The fabrication process was a combination of lithography and etching processes similar to the fabrication of the binary masks. However, it was more challenging since multiple lithography steps with precise alignment, and accurate etching time were required to achieve a practical quaternary Huffman-like mask. A schematic of the fabrication process is depicted in Fig.~\ref{fig:4level Fab Process}. A 4-inch $\textrm{SiO}_2$ wafer was coated with approximately 5 $\mu$m Ta using the sputtering machine and the same sputtering parameters as explained in Sec. \ref{sec:fabProcess_binary_masks} (see Fig.~\ref{fig:4level Fab Process} (a)). Figures~\ref{fig:4level Fab Process} (b) – (d) show the first lithography process to define level one of the quaternary Huffman-like masks. The lithography parameters which include type of the photoresist, spin-coating parameters, photoresist backing time, writing parameters as well as the development solution and development time were the same as that for the binary mask. The lithography pattern, however, was different. It contained 15 quaternary masks with five different sizes (i.e. 11, 15, 32, 43, and 86) and three different resolutions (i.e. 10 $\mu$m, 15 $\mu$m, and 20 $\mu$m). The design also included a few test patterns such as circles and bars. KLayout software was used to design the wafer. The pattern for each mask level was drawn in a separate layer to be used for each lithography step. A part of the KLayout design is shown in Fig.~\ref{fig:Huffman layout}. Figure~\ref{fig:Huffman layout}(a) is layer one (corresponding to level one) of the $15\times15$ quaternary mask with 20 $\mu$m resolution. Layer 2 (corresponding to level 2) and layer 3 (corresponding to level 3) of the same mask are illustrated in Fig.~\ref{fig:Huffman layout} (b) and (c) respectively. Figure~\ref{fig:Huffman layout} (d) is a combination of all layers, which indicate the final result. Note that level 0 has a thickness of approximately 5 $\mu$m, which is the thickness of the deposited Ta and can be seen as the background in Fig.~\ref{fig:Huffman layout} (d).

After the first lithography step, the wafer was etched (see Fig.~\ref{fig:4level Fab Process} (e)) using the RIE machine and the same recipe to deep etch Ta, as discussed in the previous section. However, the etching time was 30 minutes to provide approximately 2.2 $\mu$m Ta thickness, which was required for level one of the quaternary masks. The wafer was then washed with acetone to remove the remaining photoresist and rinsed by IPA (see Fig.~\ref{fig:4level Fab Process} (f)). At this stage the wafer was ready for the next lithography step, which is shown in Fig.~\ref{fig:4level Fab Process} (g)-(h). After spin-coating the photoresist on the etched wafer and the baking process, the patterns in layer 2 of the KLayout design (which correspond to level 2 of the quaternary Huffman-like mask) were written into the photoresist using the mask-less aligner tool. To precisely align the patterns from one layer to the other layer, alignment marks were used in all three layers such that we aligned the alignment marks in layer 2 of the design with the alignment marks in level one, which were written to the wafer in the first lithography step.

After developing the photoresist (see Fig.~\ref{fig:4level Fab Process} (h)), the wafer was etched for the second time to define level 2 of the quaternary masks as shown in Fig.~\ref{fig:4level Fab Process} (i). The etching time was 43 minutes to achieve the required Ta thickness of approximately 0.88 $\mu$m. After the etching process, the remaining photoresist was removed as shown in Fig.~\ref{fig:4level Fab Process} (j). Then, the third lithography step was performed to write the third layer (see Fig.~\ref{fig:4level Fab Process} (k)-(l)) utilising the same lithography parameters as the previous lithography steps. This was followed by the third etching process. The etching time for level 3 was 60 minutes to remove all the Ta film and reach to the $\textrm{SiO}_2$ layer as shown in Fig.~\ref{fig:4level Fab Process} (m). The last steps of the fabrication process were to remove the remaining photoresist with acetone, wash and dry the wafer with IPA and nitrogen gas respectively. The result was having 4 levels of Ta thicknesses on a $\textrm{SiO}_2$ substrate as shown in Fig.~\ref{fig:4level Fab Process} (n).

\bibliography{references}

\end{document}